\documentclass[aps,prb,reprint,superscriptaddress,nofootinbib,floatfix]{revtex4-2}
\usepackage[utf8]{inputenc}
\DeclareUnicodeCharacter{FFFC}{}
\usepackage{mathrsfs}
\usepackage{amsfonts}
\usepackage{epsfig}
\usepackage{tabularx}
\usepackage{amsmath}    % need for subequations
\usepackage{amssymb}   % defines \lesssim, etc
\usepackage{graphicx}   % need for figures
\graphicspath{{figs/}}
\usepackage{verbatim}   % useful for program listings
\usepackage{booktabs}
\usepackage{color}      % use if color is used in text
\usepackage{orcidlink}
\usepackage{tikz}
\usetikzlibrary{arrows.meta,calc,decorations.pathreplacing}
\usepackage{mathtools}
\usepackage{subcaption}
\usepackage{url}
\usepackage{hyperref}
\definecolor{nicered}{rgb}{0.5,0.,0.}
\definecolor{nicegreen}{rgb}{0.,0.5,0.}
\definecolor{niceblue}{rgb}{0.,0.,0.5}
\hypersetup{colorlinks,citecolor=nicegreen,linkcolor=nicered,urlcolor=niceblue}
\usepackage{array,multirow}
\usepackage{cprotect}
\usepackage{dcolumn}% Align table columns on decimal point
\usepackage{bm}% bold math
\usepackage{slashed}
\usepackage{physics}
\usepackage{framed}
\usepackage{algpseudocode}
\usepackage{newfloat}
\DeclareFloatingEnvironment[fileext=loa,name=Algorithm]{algorithm}

\newcolumntype{L}[1]{>{\raggedright\arraybackslash}m{#1}}

\newtheorem{definition}{Definition}[section]

\makeatother

\usepackage{enumitem}
\setlist{nolistsep}

\begin{document}
\setlength{\abovedisplayskip}{5pt}
\setlength{\belowdisplayskip}{5pt}
\title{STEC-Net: A Spatiotemporal Graph Neural Framework for Community Discovery in Dynamic Social Networks}

\author{Yingnan Xu\orcidlink{0009-0005-8296-0850}}
\email{yingnanx@mail.smu.edu}
\thanks{(Corresponding author)}
\affiliation{Department of Physics, Southern Methodist University, Dallas, Texas 75275, China\looseness=-1}
\affiliation{Department of Digital Banking, Qilu Bank, Jinan, Shandong 250014, China\looseness=-1}

\author{Shuangshuang Chu\orcidlink{0009-0004-4396-582X}}
\affiliation{Department of Digital Banking, Qilu Bank, Jinan, Shandong 250014, China\looseness=-1}

\date{\today}

\begin{abstract}
Community discovery is a central problem in the analysis of dynamic social networks. Traditional community discovery methods mainly focus on the formation and dissolution of links between nodes, and therefore often fail to capture the richer spatial structure and temporal dependency underlying network evolution. To address this limitation, \textcolor{black}{we propose STEC-Net}, a spatiotemporal graph neural framework for community discovery in dynamic social networks. STEC-Net integrates spatial structure and temporal dynamics within a unified embedding architecture. First, Graph Convolutional Networks (GCNs) are used to learn snapshot-level node representations from network topology. To adapt the spatial encoder to structural evolution, a GRU-based weight evolution mechanism is introduced to update the GCN parameters over time. Then, a second Gated Recurrent Unit (GRU) is employed to model temporal dependencies across snapshot embeddings and to learn spatiotemporal node representations. Finally, a Self-Organizing Map (SOM) is applied to the learned embeddings to cluster nodes and infer their community affiliations. Experiments on four types of dynamic networks show that STEC-Net consistently outperforms traditional community discovery methods in terms of purity, normalized mutual information, homogeneity, and completeness. These results demonstrate that STEC-Net can effectively uncover evolving community structures in dynamic social networks.
\end{abstract}


\maketitle

\section{Introduction}\label{sec:intro}

Social networks have become an indispensable part of modern life, with continuous growth in both scale and structural complexity. Dynamic social networks are composed of a large number of user nodes together with evolving interaction patterns that change over time. These networks contain rich structural information and temporal regularities whose effective analysis is essential for understanding complex social behavior.

Community discovery\citep{kang2024overlapping}, as one of the core tasks in social network analysis, aims to identify groups of nodes that are densely connected internally while remaining relatively independent from other groups. The applications of community discovery are broad and practically significant. In targeted marketing and personalization, community structure reveals common interests and behavioral preferences\citep{Paliouras}, thereby supporting recommendation systems and precision advertising. In epidemiology and public health\citep{10.1093/ije/dyh010,rostami2023}, communities may correspond to populations with elevated transmission risk, which is important for intervention design and disease containment. In security and fraud detection, community discovery can help reveal suspicious groups in financial or communication networks\citep{6957290,andre2012community,waskiewicz2012friend}, thereby improving the identification of fraudulent activities, cyber threats, and other abnormal behaviors. In social science and behavioral analysis\citep{dhawan2021community,1561484}, evolving communities provide insight into social influence, group interaction, and the formation of collective behavior. These applications demonstrate the importance of accurate community discovery in dynamic networks.

In static social networks, many community discovery algorithms\citep{zhang2024local} have achieved useful results. However, the temporal variability of dynamic social networks creates substantial challenges for community discovery\citep{he2024dynamic,wu2024review}. Traditional static methods\citep{zhang2020progress} usually process the network at a single time point and therefore cannot adequately describe the continuity, transition, and evolution of community structure across time. As a result, they often fail to reflect the true dynamic organization of the network.

Graph embedding techniques\citep{yuan2022overview} have recently provided an important representation-learning perspective for network analysis. By mapping nodes into low-dimensional latent spaces, graph embedding methods can preserve structural and semantic information more effectively than many conventional approaches. Nevertheless, most existing graph embedding methods are still designed primarily for static networks, and they remain limited in their ability to jointly model spatial structure and temporal dependency in dynamic social networks\citep{duan2021review}. This limitation becomes particularly pronounced when the network exhibits substantial structural evolution, nonstationary interactions, and changing community boundaries over time.

\textcolor{black}{
In parallel with the development of graph embedding methods, attention-based spatiotemporal graph models have shown strong representation capacity for evolving graph data. Representative methods such as ASTGCN, MSTGCN, and ST-GAT employ adaptive weighting mechanisms to capture nonuniform spatial and temporal dependencies\citep{guo2019attention,jia2021multi,song2022st}. These models enrich the methodological landscape for spatiotemporal graph learning and therefore provide meaningful benchmarks for evaluating representation quality. However, they are not specifically designed for dynamic community discovery, and their attention mechanisms often introduce additional parameter and computational overhead. At the clustering stage, lightweight methods such as NWFCM improve partition quality by incorporating node importance into the clustering process\citep{huang2024community}, yet they do not provide a unified spatiotemporal representation framework for evolving networks.}

\textcolor{black}{
To address these limitations, this paper proposes STEC-Net, a dynamic community discovery framework based on spatiotemporal graph embedding. The proposed method jointly models snapshot-level topology and temporal evolution within a unified architecture. Specifically, Graph Convolutional Networks are used to learn structural representations for each snapshot, a GRU-based weight evolution mechanism is introduced to adapt the spatial encoder to network evolution, and a second GRU is employed to capture temporal dependency across snapshots. On this basis, a Self-Organizing Map is used to infer community affiliations from the learned spatiotemporal embeddings. Through this design, STEC-Net strengthens the integration of spatial representation, temporal modeling, and adaptive clustering for dynamic community discovery.}

\textcolor{black}{
The main contributions of this paper can be summarized as follows. This paper develops a unified spatiotemporal graph neural framework for dynamic community discovery by integrating graph convolution, recurrent parameter evolution, and temporal representation learning within a single architecture. It further incorporates Self-Organizing Map clustering on top of the learned embeddings, thereby enabling adaptive community identification without requiring the number of communities to be specified in advance. Extensive experiments on synthetic and real dynamic networks, together with comparisons against classical community discovery methods, graph embedding baselines, and recent attention-based approaches, confirm the effectiveness, robustness, and computational efficiency of STEC-Net.}

This paper is organized as follows. In Sect.~\ref{sec:intro}, we introduce the research background and motivation. In Sect.~\ref{sec:works}, we review the progress and limitations of related studies. In Sect.~\ref{sec:def}, we present the necessary definitions and the proposed STEC-Net framework. Experimental results and comparative analyses are reported in Sect.~\ref{sec:results}. Finally, Sect.~\ref{sec:conc} concludes the paper.

\section{Related Work}\label{sec:works}

Research on community discovery in dynamic social networks has attracted sustained attention because real communities are not static but continuously evolve with changing interaction patterns\citep{he2024dynamic,wu2024review}. Existing studies may be broadly understood from three perspectives: traditional community discovery methods, graph embedding methods, and recent spatiotemporal graph neural approaches. Each line of work has contributed important ideas, yet important limitations remain for dynamic community discovery.

The first line of research extends conventional community discovery methods from static networks to evolving settings. Static community discovery algorithms\citep{zhang2024local,zhang2020progress} have achieved useful results in identifying densely connected groups in fixed network structures. However, when directly applied to dynamic networks, such methods generally process snapshots independently and therefore cannot preserve temporal continuity in community evolution. Subsequent dynamic community discovery studies\citep{he2024dynamic,wu2024review} improve upon this limitation by incorporating historical information, evolution events, or temporal update strategies. Although these methods enhance temporal awareness, many of them still emphasize local or short-horizon changes and therefore may struggle to model long-range dependency, strong structural variation, and the joint interaction between spatial topology and temporal evolution.

The second line of research concerns graph embedding for network representation learning. Graph embedding methods map nodes into low-dimensional latent spaces while preserving important structural information\citep{yuan2022overview}. This idea provides a more flexible basis for downstream tasks such as node classification, link prediction, and community discovery. For dynamic networks, however, the essential challenge is not only to obtain high-quality snapshot-level representations, but also to couple these representations with temporal evolution in a coherent manner\citep{duan2021review}. Many existing embedding methods still focus primarily on static topology or on limited temporal updates, which constrains their ability to characterize evolving communities in networks with nonstationary interaction patterns and changing structural boundaries.

\textcolor{black}{
More recently, spatiotemporal graph neural networks have further advanced representation learning on dynamic graphs. Attention-based models such as ASTGCN, MSTGCN, and ST-GAT enhance spatiotemporal modeling by assigning adaptive importance weights to graph neighborhoods and temporal contexts\citep{guo2019attention,jia2021multi,song2022st}. These methods demonstrate that attention mechanisms can improve the extraction of informative spatial and temporal dependencies, and they therefore provide strong comparative baselines for dynamic graph representation learning. Nevertheless, they are primarily general spatiotemporal learning architectures rather than dedicated community discovery frameworks, and their attention modules may introduce substantial computational and parameter overhead when multi-snapshot modeling is required. In addition, lightweight clustering-oriented methods such as NWFCM improve partition quality by incorporating node importance into the clustering process\citep{huang2024community}, but they do not directly solve the problem of unified spatiotemporal representation learning for dynamic community discovery.}

In summary, existing studies have established important foundations for dynamic community analysis, graph embedding, and spatiotemporal graph learning. However, there remains a clear need for a framework that can simultaneously capture snapshot-level structural information, adapt to temporal evolution, support adaptive community partitioning, and maintain favorable computational efficiency. STEC-Net is proposed to address this need by integrating graph convolution, recurrent temporal modeling, and Self-Organizing Map clustering within a unified dynamic community discovery framework.

\section{Dynamic Community Discovery Algorithm}\label{sec:def}

In this section, we first introduce the notation and fundamental concepts used throughout the paper, and then present the proposed dynamic community discovery framework.

\subsection{Definitions}

\begin{definition}[Dynamic Network]
Let \(V=\{v_1,v_2,\dots,v_n\}\) denote a fixed set of nodes, and let \(E \subseteq V \times V\) denote the set of all possible edges. At time step \(t\), the network snapshot is defined as
\[
g_t=(V,E_t),
\]
where \(E_t \subseteq E\) may vary over time, whereas the node set \(V\) remains unchanged. The dynamic network over an observation horizon of length \(T\) is therefore represented as
\[
G=\{g_1,g_2,\dots,g_T\}.
\]

Each snapshot \(g_t\) is associated with an adjacency matrix \(A_t \in \mathbb{R}^{n \times n}\), where \(A_t(i,j)\) denotes the edge weight between nodes \(v_i\) and \(v_j\) at time \(t\). Unless otherwise specified, all graphs considered in this paper are undirected and unweighted, so that \(A_t=A_t^\top\) and \(A_t(i,j)\in\{0,1\}\).
\end{definition}

\begin{definition}[Dynamic Community Structure]
Let
\[
C_t=\{c_t^1,c_t^2,\dots,c_t^{p_t}\}
\]
denote the set of communities in snapshot \(g_t\), where each \(c_t^k \subseteq V\) is the set of nodes belonging to the \(k\)-th community at time \(t\), and \(p_t\) is the number of communities at time \(t\). The community structure over the full dynamic network is denoted by
\[
C=\{C_1,C_2,\dots,C_T\}.
\]
\end{definition}

\begin{definition}[Dynamic Node Embedding]
Let \(\mathcal{E}\) denote the overall embedding function that maps node \(v_i\) at time \(t\) to a \(d\)-dimensional latent representation:
\[
\mathcal{E}:(v_i,t)\mapsto y_i^t \in \mathbb{R}^d,
\qquad
1\le i \le n,\;\; 1\le t \le T,\;\; d \ll n.
\]
The final spatiotemporal embedding matrix at time \(t\) is
\[
Y_t=[y_1^t,y_2^t,\dots,y_n^t]^\top \in \mathbb{R}^{n \times d},
\]
and the full embedding sequence is denoted by
\[
Y=\{Y_1,Y_2,\dots,Y_T\}.
\]
\end{definition}

\begin{definition}[Normalized Adjacency Matrix]
For each snapshot \(g_t\), the adjacency matrix is defined as
\[
A_t(i,j)=
\begin{cases}
1, & \text{if } (v_i,v_j)\in E_t,\\
0, & \text{otherwise},
\end{cases}
\qquad \forall v_i,v_j \in V.
\]
To perform graph convolution, self-loops are added to the graph:
\[
\tilde{A}_t=A_t+\mathbb{I},
\]
where \(\mathbb{I}\in\mathbb{R}^{n\times n}\) is the identity matrix. Let
\[
\tilde{D}_t=\operatorname{diag}\!\left(\sum_{j=1}^n \tilde{A}_t(i,j)\right)
\]
be the degree matrix of \(\tilde{A}_t\). The symmetrically normalized adjacency matrix is then given by
\[
\hat{A}_t=\tilde{D}_t^{-1/2}\tilde{A}_t\tilde{D}_t^{-1/2}.
\]
\end{definition}

\begin{table}[ht]
\centering
\begin{tabular}{ll}
\toprule
\textbf{Symbol} & \textbf{Description} \\
\midrule
\(G\) & Dynamic network \( \{g_t\}_{t=1}^T \) \\
\(g_t\) & Network snapshot at time \(t\) \\
\(V\) & Node set \\
\(E_t\) & Edge set at time \(t\) \\
\(A_t\) & Adjacency matrix of \(g_t\) \\
\(\tilde{A}_t\) & Adjacency matrix with self-loops \\
\(\hat{A}_t\) & Symmetrically normalized adjacency matrix \\
\(T\) & Number of time steps \\
\(L\) & Number of GCN layers \\
\(n\) & Number of nodes \\
\(d_0\) & Input feature dimension \\
\(d_l\) & Hidden dimension of the \(l\)-th GCN layer \\
\(d\) & Dimension of the final embedding \\
\(\mathcal{E}\) & Overall embedding function \\
\(X_t\) & Input node feature matrix at time \(t\) \\
\(H_t^{(l)}\) & Hidden node representation at GCN layer \(l\) and time \(t\) \\
\(F_t\) & Snapshot-level spatial embedding at time \(t\) \\
\(Y_t\) & Final spatiotemporal embedding at time \(t\) \\
\(W_t^{(l)}\) & GCN weight matrix of layer \(l\) at time \(t\) \\
\(Q_t^{(l)}\) & Summary matrix used to update \(W_t^{(l)}\) \\
\(\mathcal{S}^{(l)}(\cdot)\) & Learnable summarization operator at layer \(l\) \\
\(C_t\) & Community set at time \(t\) \\
\(C\) & Dynamic community sequence \( \{C_t\}_{t=1}^T \) \\
\(\Gamma^{(l)}(\cdot)\) & Graph convolution operator at layer \(l\) \\
\(\Phi^{(l)}(\cdot)\) & GRU-based weight evolution operator at layer \(l\) \\
\(\Psi(\cdot)\) & Temporal GRU operator \\
\(\mathcal{P}_t\) & Set of positive node pairs at time \(t\) \\
\(\mathcal{N}_t(i)\) & Set of negative nodes for anchor node \(v_i\) at time \(t\) \\
\(P_n^t(v)\) & Negative-sampling distribution at time \(t\) \\
\(\mathcal{L}\) & Training loss function \\
\(m\) & Margin hyperparameter in the ranking loss \\
\(Q\) & Number of negative samples per positive pair \\
\bottomrule
\end{tabular}
\caption{Notation used throughout the paper.}
\label{tab:symbols}
\end{table}

We summarize the principal symbols used in the proposed framework in Tab.~\ref{tab:symbols}.

\subsection{The STEC-Net Algorithm}

\begin{figure}[ht]
\centering
\includegraphics[width=\linewidth]{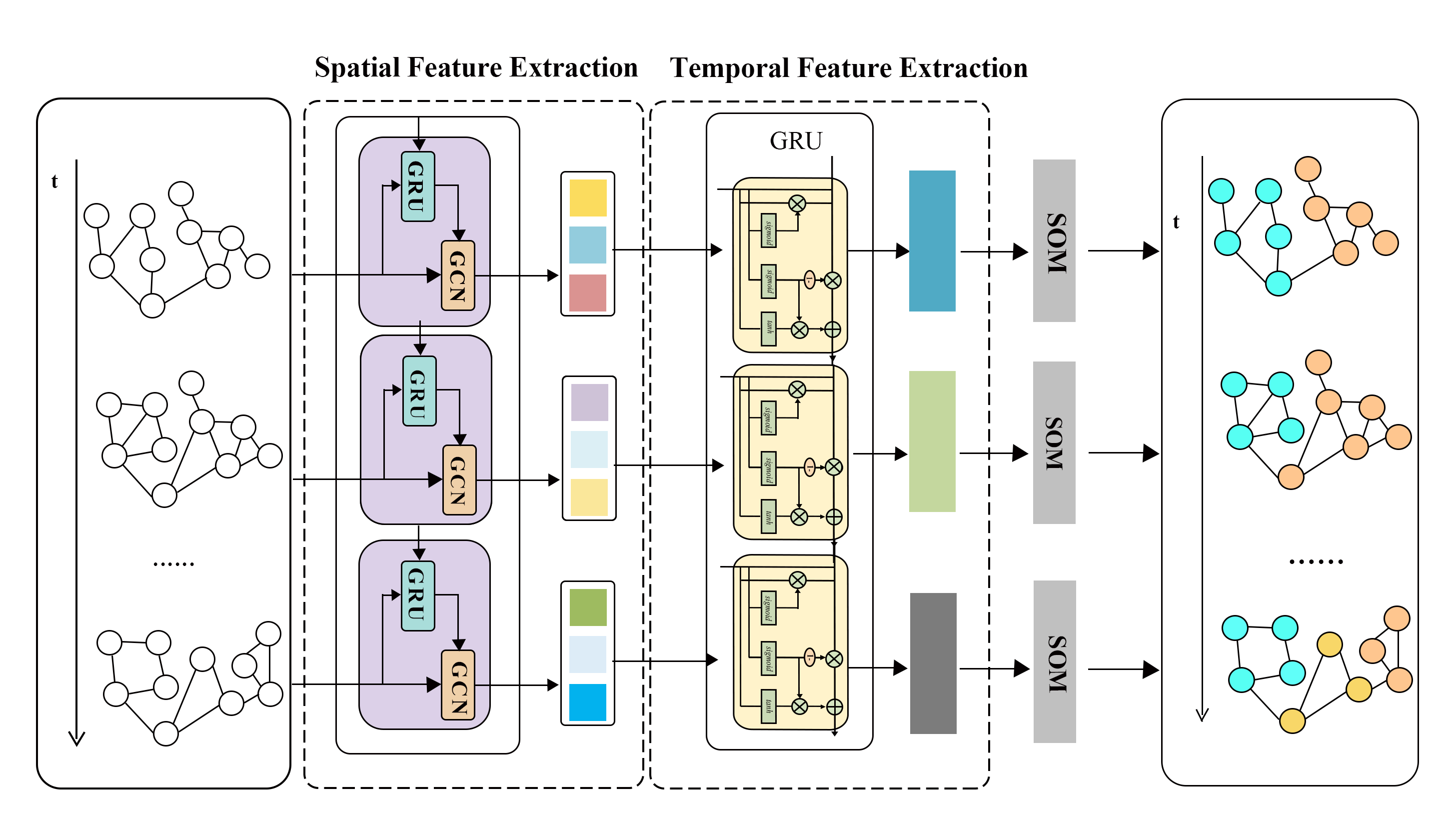}
\caption{Overview of the proposed STEC-Net framework. The model first learns snapshot-level spatial node representations by GCNs, then updates GCN parameters over time through a GRU-based weight evolution mechanism, further captures temporal dependency by a second GRU operating on spatial embeddings, and finally identifies dynamic communities by SOM.}
\label{fig:1}
\end{figure}

The proposed STEC-Net framework, illustrated in Fig.~\ref{fig:1}, consists of three tightly coupled components: spatial feature extraction, temporal feature extraction, and community discovery. First, a graph convolutional network (GCN) is employed to learn snapshot-wise node representations from the network topology \citep{gcn}. Second, in order to adapt the spatial encoder to structural evolution, the GCN weight matrices are updated over time by a GRU-based parameter evolution mechanism \citep{duan2021review,pareja2020evolvegcn}. Third, the resulting spatial embeddings \(F_t\) are fed into a temporal GRU to capture temporal dependency and long-range evolution patterns across network snapshots \citep{gru}. Finally, a Self-Organizing Map (SOM) is applied to the learned spatiotemporal embeddings \(Y_t\) to identify the dynamic community structure \citep{som,som2}.

\subsection{Spatial Feature Extraction}

The purpose of the spatial feature extraction stage is to learn node representations that preserve the structural information of each snapshot. This stage contains two components: snapshot-level graph embedding and temporal evolution of GCN parameters.

\subsubsection{Spatial Embedding Learning}

\textcolor{black}{Let \(X_t \in \mathbb{R}^{n \times d_0}\) denote the input node feature matrix associated with snapshot \(g_t\), and define
\[
H_t^{(0)} = X_t.
\]
Because explicit time-varying node attributes are unavailable in the dynamic networks considered in this paper, \(X_t\) is treated as a latent snapshot-specific input representation and is randomly initialized before model training.}

For the \(l\)-th GCN layer (\(l=1,2,\dots,L\)), the hidden representation is computed as
\begin{equation}
H_t^{(l)}
=
\Gamma^{(l)}\!\left(H_t^{(l-1)},\hat{A}_t,W_t^{(l)}\right)
=
\sigma\!\left(\hat{A}_t H_t^{(l-1)} W_t^{(l)}\right),
\label{eq:gcn}
\end{equation}
where \(H_t^{(l-1)} \in \mathbb{R}^{n \times d_{l-1}}\) is the input representation of layer \(l\), \(W_t^{(l)} \in \mathbb{R}^{d_{l-1} \times d_l}\) is the learnable weight matrix at time \(t\), and \(\sigma(\cdot)\) is the nonlinear activation function. In this paper, LeakyReLU is adopted. Equation~\eqref{eq:gcn} follows the standard first-order GCN formulation with symmetric adjacency normalization \citep{gcn}.

The output of the final GCN layer is taken as the spatial embedding of snapshot \(g_t\):
\begin{equation}
F_t = H_t^{(L)} \in \mathbb{R}^{n \times d}.
\label{eq:spatial_embedding}
\end{equation}
Accordingly, the sequence of snapshot-level spatial embeddings is denoted by
\[
F=\{F_1,F_2,\dots,F_T\}.
\]

The normalized adjacency matrix \(\hat{A}_t\) in Eq.~\eqref{eq:gcn} is constructed as follows:
\begin{equation}
\tilde{A}_t = A_t + \mathbb{I},
\label{eq:adj_aug}
\end{equation}
\begin{equation}
[\tilde{D}_t]_{ii} = \sum_{j=1}^n \tilde{A}_t(i,j), \qquad i=1,\dots,n,
\label{eq:deg}
\end{equation}
\begin{equation}
\hat{A}_t = \tilde{D}_t^{-1/2}\tilde{A}_t\tilde{D}_t^{-1/2}.
\label{eq:adj_norm}
\end{equation}
This normalization stabilizes neighborhood aggregation and reduces the influence of node-degree variation \citep{gcn}.

\subsubsection{Weight Parameter Evolution}

\textcolor{black}{
Since the network topology changes over time, using a single fixed set of GCN parameters for all snapshots is suboptimal. Following the idea of EvolveGCN-style dynamic parameter updating \citep{duan2021review,pareja2020evolvegcn}, we employ a GRU-based mechanism to evolve the GCN weights over time.}

\textcolor{black}{
For dimensional compatibility, the input to the recurrent update of the \(l\)-th GCN layer is first summarized into a matrix
\[
Q_t^{(l)}=\mathcal{S}^{(l)}\!\left(H_t^{(l-1)}\right) \in \mathbb{R}^{d_{l-1}\times d_l},
\]
where \(\mathcal{S}^{(l)}(\cdot)\) denotes a learnable summarization operator. The weight matrix \(W_t^{(l)}\) is then updated from the previous weight \(W_{t-1}^{(l)}\) through a GRU cell:
\begin{equation}
Z_t^{(l)}=
\sigma\!\left(Q_t^{(l)}U_Z^{(l)} + W_{t-1}^{(l)}V_Z^{(l)} + B_Z^{(l)}\right),
\label{eq:gru_weight_z}
\end{equation}
\begin{equation}
R_t^{(l)}=
\sigma\!\left(Q_t^{(l)}U_R^{(l)} + W_{t-1}^{(l)}V_R^{(l)} + B_R^{(l)}\right),
\label{eq:gru_weight_r}
\end{equation}
\begin{equation}
\widetilde{W}_t^{(l)}=
\tanh\!\left(Q_t^{(l)}U_W^{(l)} + \left(R_t^{(l)}\circ W_{t-1}^{(l)}\right)V_W^{(l)} + B_W^{(l)}\right),
\label{eq:gru_weight_candidate}
\end{equation}
\begin{equation}
W_t^{(l)}=
\left(1-Z_t^{(l)}\right)\circ W_{t-1}^{(l)} + Z_t^{(l)}\circ \widetilde{W}_t^{(l)},
\label{eq:gru_weight_update}
\end{equation}
where \(Z_t^{(l)}\) and \(R_t^{(l)}\) are the update and reset gates, respectively, \(U_{\ast}^{(l)}\) and \(V_{\ast}^{(l)}\) are trainable parameter matrices, \(B_{\ast}^{(l)}\) are bias terms, and \(\circ\) denotes the Hadamard product. }

Therefore, for each layer \(l\), the spatial feature extraction process at time \(t\) can be written compactly as
\begin{widetext}
\begin{equation}
W_t^{(l)} = \Phi^{(l)}\!\left(Q_t^{(l)},W_{t-1}^{(l)}\right),
\qquad
H_t^{(l)} = \Gamma^{(l)}\!\left(H_t^{(l-1)},\hat{A}_t,W_t^{(l)}\right).
\label{eq:spatial_compact}
\end{equation}
\end{widetext}
In this way, the GCN captures snapshot-level structural information, while the GRU-based parameter evolution allows the spatial encoder to adapt continuously to the dynamic network.

\begin{figure}[t]
\centering
\includegraphics[width=\linewidth]{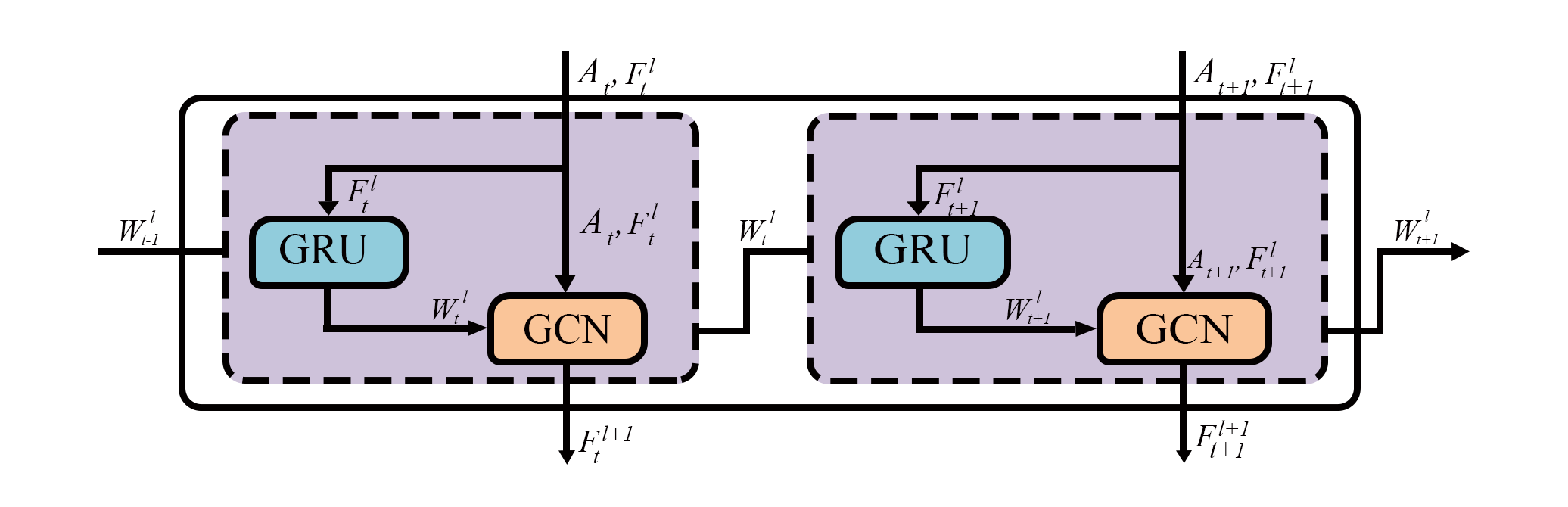}
\caption{Spatial feature extraction stage of STEC-Net. GCNs learn snapshot-level node representations \(F_t\), while the GCN parameters are evolved across time by a GRU-based update mechanism.}
\label{fig:2}
\end{figure}

The spatial feature extraction stage of STEC-Net is illustrated in Fig.~\ref{fig:2}.

\subsection{Temporal Feature Extraction}

Although the spatial encoder captures the structural characteristics of each individual snapshot, it does not by itself model temporal dependency across snapshots. In dynamic networks, node interactions may reappear after a long interval, and such long-range dependency is difficult to capture by purely snapshot-based models. To address this issue, we apply a temporal GRU to the sequence of spatial embeddings \(\{F_t\}_{t=1}^T\) \citep{gru}.

It is important to note that this GRU is distinct from the GRU used in the previous subsection for GCN weight evolution. The former updates model parameters, whereas the present GRU updates node representations over time.

\textcolor{black}{
Let \(Y_{t-1} \in \mathbb{R}^{n \times d}\) denote the hidden state inherited from the previous time step, with \(Y_0\) initialized as a zero matrix. Taking the spatial embedding \(F_t\) as the current input, the reset gate and update gate are computed as
\begin{equation}
r_t=\sigma\!\left(F_t W_r + Y_{t-1} U_r + \mathbf{1} b_r^\top\right),
\label{eq:gru_time_r}
\end{equation}
\begin{equation}
z_t=\sigma\!\left(F_t W_z + Y_{t-1} U_z + \mathbf{1} b_z^\top\right),
\label{eq:gru_time_z}
\end{equation}
where \(W_r,W_z,U_r,U_z \in \mathbb{R}^{d \times d}\), \(b_r,b_z \in \mathbb{R}^{d}\), and \(\mathbf{1}\in\mathbb{R}^{n}\) is the all-ones vector used to broadcast the bias terms across nodes. The candidate hidden state is then obtained by
\begin{equation}
\widetilde{Y}_t=
\tanh\!\left(F_t W_h + \left(r_t \circ Y_{t-1}\right) U_h + \mathbf{1} b_h^\top\right),
\label{eq:gru_time_candidate}
\end{equation}
and the final spatiotemporal embedding is updated as
\begin{equation}
Y_t=
\left(1-z_t\right)\circ Y_{t-1} + z_t \circ \widetilde{Y}_t.
\label{eq:gru_time_update}
\end{equation}
Here \(W_h,U_h \in \mathbb{R}^{d \times d}\) and \(b_h \in \mathbb{R}^{d}\).}

\begin{figure}[t]
\centering
\includegraphics[width=0.6\linewidth]{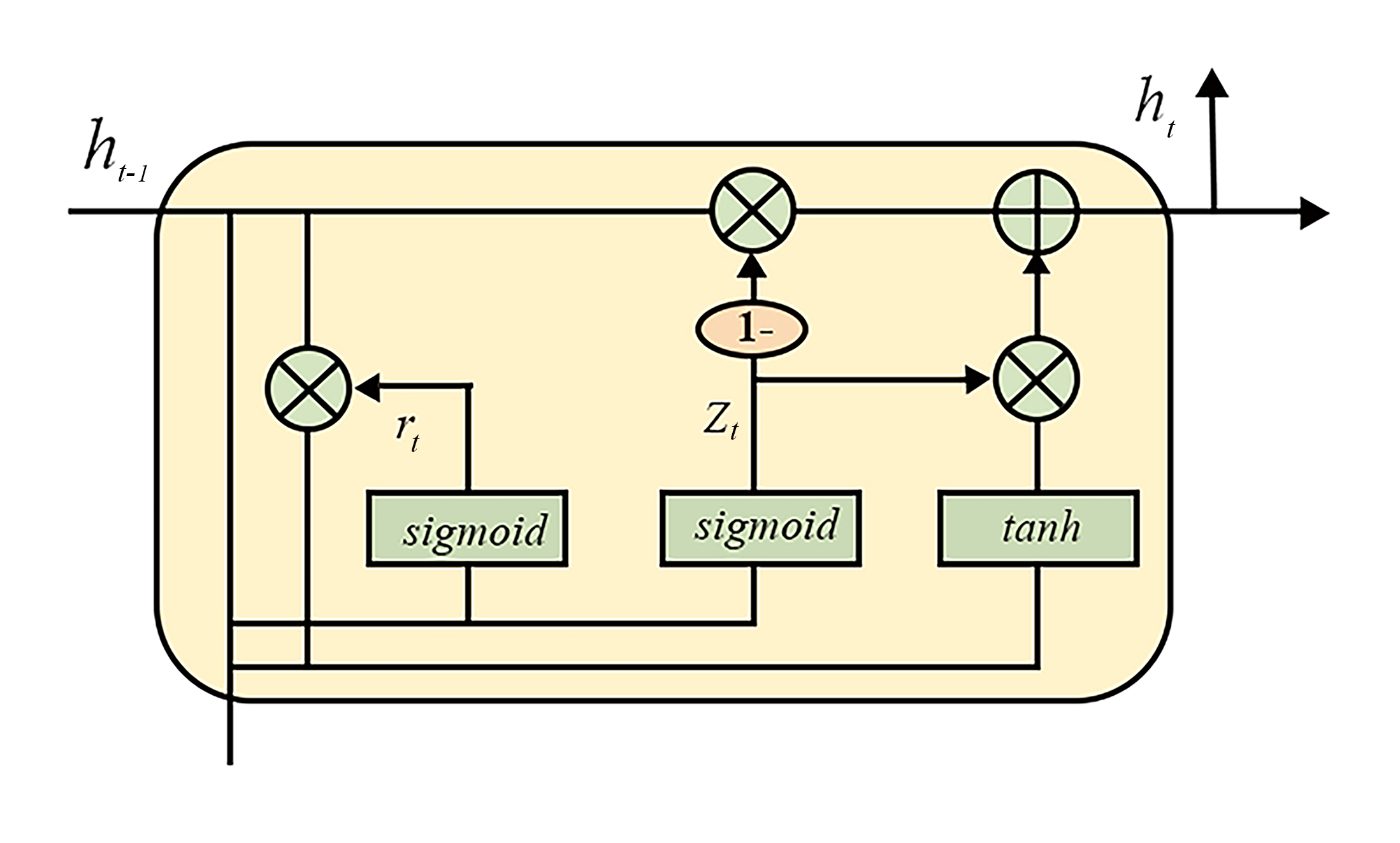}
\caption{GRU architecture used in the temporal feature extraction module of STEC-Net, where the spatial embedding \(F_t\) is transformed into the final spatiotemporal embedding \(Y_t\).}
\label{fig:3}
\end{figure}

The GRU structure used in the temporal feature extraction module is shown in Fig.~\ref{fig:3}. By combining the current spatial embedding \(F_t\) with the historical hidden state \(Y_{t-1}\), the model captures both short-term evolution and long-range temporal dependency in the dynamic network.

\subsection{Community Detection}

The proposed model learns node representations in an unsupervised manner. To ensure that the learned embeddings preserve graph topology, we adopt a pairwise distance-based objective: nodes connected by an edge should be embedded close to one another, whereas unconnected nodes should be separated in the latent space.

For snapshot \(g_t\), let
\[
\mathcal{P}_t=\{(i,j)\mid (v_i,v_j)\in E_t\}
\]
denote the set of positive node pairs, and let
\[
\mathcal{N}_t(i)=\{u\in V \mid u\neq i,\; (v_i,v_u)\notin E_t\}
\]
denote the set of negative nodes associated with anchor node \(v_i\). A margin-based ranking loss can then be defined as
\begin{equation}
\mathcal{L}
=
\sum_{t=1}^{T}
\sum_{(i,j)\in \mathcal{P}_t}
\sum_{u\in \mathcal{N}_t(i)}
\left[
m+\|y_i^t-y_j^t\|_2^2-\|y_i^t-y_u^t\|_2^2
\right]_+,
\label{eq:loss_full}
\end{equation}
where \([\cdot]_+=\max(\cdot,0)\) is the hinge loss, and \(m>0\) is a margin hyperparameter controlling the desired separation between positive and negative pairs.

For large-scale graphs, enumerating all negative pairs is computationally prohibitive. Therefore, following the negative sampling strategy adopted in related dynamic network studies \citep{duan2021review,yang2023social} and the standard negative-sampling formulation \citep{mikolov2013distributed}, we sample a small number of unconnected nodes for each positive pair. Let \(Q\) denote the number of negative samples, and let \(P_n^t(v)\) be the noise distribution at time \(t\), defined proportionally to node degree as \(P_n^t(v)\propto (d_v^t)^{3/4}\). The practical training objective is written as
\begin{widetext}
\begin{equation}
\mathcal{L}
=
\sum_{t=1}^{T}
\sum_{(i,j)\in \mathcal{P}_t}
\left(
\sum_{q=1}^{Q}
\mathbb{E}_{v_q \sim P_n^t(v)}
\left[
m+\|y_i^t-y_j^t\|_2^2-\|y_i^t-y_{v_q}^t\|_2^2
\right]_+
\right),
\label{eq:loss_ns}
\end{equation}
\end{widetext}
where \(Q\) is the number of negative samples drawn for each positive pair.

After learning the spatiotemporal embeddings \(Y_t\), we apply the Self-Organizing Map (SOM) to discover latent community structure \citep{som,som2}. Accordingly, the community structure at time \(t\) is obtained as
\[
C_t=\mathrm{SOM}(Y_t),
\]
and the full dynamic community sequence is
\[
C=\{C_1,C_2,\dots,C_T\}.
\]

For clarity, the complete STEC-Net procedure is summarized in Algorithm~\ref{alg:STEC-Net}. The formulation is written for a general \(L\)-layer GCN; in our implementation, \(L=2\).

\begin{algorithm}[ht]
\caption{Dynamic Community Discovery Method Based on Spatiotemporal Graph Embedding}
\label{alg:STEC-Net}

\noindent\fbox{%
\begin{minipage}{0.96\linewidth}
\begin{algorithmic}[1]
\Require Dynamic network \(G=\{g_t\}_{t=1}^{T}\); number of GCN layers \(L\)
\Ensure Dynamic community sequence \(C=\{C_t\}_{t=1}^{T}\)

\State Initialize \(\{W_0^{(l)}\}_{l=1}^{L}\)
\State Initialize temporal hidden state \(Y_0\)
\State Initialize \(\{X_t\}_{t=1}^{T}\), where \(X_t \in \mathbb{R}^{n \times d_0}\)

\For{$t=1$ to $T$}
    \State Construct adjacency matrix \(A_t\)
    \State Compute \(\tilde{A}_t = A_t + \mathbb{I}\), \(\tilde{D}_t\), and \(\hat{A}_t = \tilde{D}_t^{-1/2}\tilde{A}_t\tilde{D}_t^{-1/2}\)
    \State Set \(H_t^{(0)} \gets X_t\)

    \For{$l=1$ to $L$}
        \State Compute summary matrix \(Q_t^{(l)} \gets \mathcal{S}^{(l)}(H_t^{(l-1)})\)
        \State Update GCN weights \(W_t^{(l)} \gets \Phi^{(l)}(Q_t^{(l)}, W_{t-1}^{(l)})\)
        \State Compute node representation \(H_t^{(l)} \gets \Gamma^{(l)}(H_t^{(l-1)}, \hat{A}_t, W_t^{(l)})\)
    \EndFor

    \State Obtain spatial embedding \(F_t \gets H_t^{(L)}\)
    \State Obtain spatiotemporal embedding \(Y_t \gets \Psi(F_t, Y_{t-1})\)
    \State Obtain community assignment \(C_t \gets \mathrm{SOM}(Y_t)\)
    \State Append \(C_t\) to \(C\)
\EndFor

\Return \(C\)
\end{algorithmic}
\end{minipage}%
}

\end{algorithm}

\section{Results and Analysis}\label{sec:results}
\subsection{Experiment Setup}

All algorithms in this study are implemented in \texttt{Python 3.10}, with the GCN and GRU components developed using \texttt{PyTorch 2.4.1}. Experiments are conducted on a Hygon C86 32-core CPU machine with a Kylin 10 Linux-based operating system, equipped with 512 GB of memory and 8 Hygon DCU GPUs.

The datasets used in this study consist of time-snapshot dynamic networks, including both synthetic and real-world network datasets. For the synthetic data, we adopt the widely used LFR (Lancichinetti--Fortunato--Radicchi) benchmark model\cite{lfr}, which generates graphs with built-in community structure. In this simulation: $N = 1000$ specifies the total number of nodes in the graph; \( s = 9 \) defines the number of time snapshots (i.e., network evolution stages); and \( \mu \in [0.1, 0.8] \) is the mixing parameter, which controls the community strength: lower values of \( \mu \) indicate clearer communities (most edges remain within communities), while higher values increase the inter-community connections, making communities harder to detect. By varying \( \mu \), we generate eight synthetic dynamic network datasets labeled from LFR1 to LFR8. These provide different levels of community detectability over time.

To evaluate our STEC-Net model, we also apply it to four real-world dynamic networks of varying size and complexity: PS network representing face-to-face interactions in educational settings~\cite{yang2012defining}, CSMN network capturing online social interactions in college environments~\cite{cornfield2023buzz}, DBLP collaboration network from computer science publications~\cite{yang2012defining}, and Brain networks derived from neuroimaging data~\cite{brain,MJNMFGAT}. Cluster indicators were used as evaluation criteria in the experiment,  mainly including purity, normalized mutual information, homogeneity and completeness.

	The GCN is configured with $L=2$ layers, input feature dimension $d_0=64$, and output embedding dimension $d=64$. 
	All learnable parameters are optimized using the Adam optimizer with a learning rate of $0.001$ over $200$ epochs. 
	In the ranking loss, the margin parameter is set to $m=1.0$ and the number of negative samples per positive pair to $Q=5$. 
	The SOM is configured with a $10\times10$ competitive unit grid, an initial neighborhood radius $\sigma_0=3.0$, and an initial learning rate $\eta_0=0.5$, both decaying linearly during training. All experiments were repeated five times with different random seeds, and the reported results are mean $\pm$ standard deviation. The standard deviation is used to reflect the stability of each method under random initialization and stochastic training.

\subsection{Evaluation Metrics}

The experimental evaluation employs four clustering quality metrics to assess community discovery performance:

\textbf{Purity:} Measures the proportion of correctly classified nodes within each cluster, calculated as
\begin{equation}
\text{Purity} = \frac{1}{N} \sum_{k=1}^K \max_j |C_k \cap T_j|
\end{equation}
where $C_k$ represents the $k$-th discovered community and $T_j$ represents the $j$-th true community. Higher purity values indicate better clustering quality.

\textbf{Normalized Mutual Information (NMI):} Quantifies the agreement between discovered and ground-truth communities, defined as
\begin{equation}
\text{NMI} = \frac{2 \cdot I(C,T)}{H(C) + H(T)}
\end{equation}
where $I(C,T)$ is the mutual information and $H(\cdot)$ denotes entropy. NMI ranges from 0 to 1, with values closer to 1 indicating better performance.

\textbf{Homogeneity:} Measures whether each cluster contains only members of a single class, ensuring internal consistency within discovered communities.

\textbf{Completeness:} Evaluates whether all members of a given class are assigned to the same cluster, assessing the ability to group related nodes together.

\subsection{Datasets}

In this experiment, synthetic and real networks are used to comprehensively evaluate algorithm performance. The LFR benchmark networks provide controllable community structures for systematic testing. Real datasets include: PS, CSMN, Brain and DBLP. The summary of these datasets structures can be found in Tab.~\ref{tab:datasets}

\subsubsection{Synthetic Datasets}

We used the LFR benchmark to generate dynamic artificial datasets. The LFR benchmark is widely recognized as a standard for evaluating community detection algorithms because it can generate networks with controllable community structure and realistic properties. The following parameters were used:

\begin{itemize}
\item $N=1000$: Total number of nodes in the network
\item $s=9$: Number of time snapshots in the dynamic sequence
\item $\mu \in [0.1,0.8]$: Mixing parameter that controls network complexity
\item Average degree: $\langle k \rangle = 15$
\item Maximum degree: $k_{max} = 30$
\item Community size range: $[10, 50]$ nodes per community
\item Power-law exponent for degree distribution: $\gamma = 2.5$
\item Power-law exponent for community size distribution: $\beta = 1.5$
\end{itemize}

The mixing parameter $\mu$ determines the fraction of edges that connect nodes to communities other than their own, where smaller $\mu$ values indicate clearer community structure and larger $\mu$ values represent more ambiguous community boundaries. By varying only the value of $\mu$, we obtained 8 dynamic artificial datasets denoted as LFR1 ($\mu=0.1$), LFR2 ($\mu=0.2$), LFR3 ($\mu=0.3$), LFR4 ($\mu=0.4$), LFR5 ($\mu=0.5$), LFR6 ($\mu=0.6$), LFR7 ($\mu=0.7$), and LFR8 ($\mu=0.8$), corresponding to different levels of community structure clarity from very clear to highly ambiguous.

\subsubsection{Real-world Datasets}

To validate the performance of STEC-Net on real-world scenarios, we employed four types of dynamic networks with varying scales and characteristics:

\textbf{PS (Primary School) Dataset:} This dataset records the face-to-face interactions between students and teachers in a primary school in Lyon, France, collected over two days using wearable sensors. The network contains 242 nodes (students and teachers) and evolves over 125 time steps with 20-second intervals. The dataset captures the natural formation and dissolution of social groups in an educational environment, making it ideal for studying temporal community dynamics.

\textbf{CSMN (College Social Message Network) Dataset:} This dataset represents online social interactions among college students over a semester period. It contains 1,899 nodes representing students and 59,835 temporal edges representing various types of interactions including messages, wall posts, and friendship formations. The network evolves over 193 time steps, each representing one day. This dataset is characterized by periodic patterns corresponding to academic schedules and social events.

\textbf{DBLP Dataset:} This is a collaboration network derived from the DBLP computer science bibliography database, focusing on publications from 2000 to 2005. The network contains 12,334 authors as nodes and co-authorship relationships as edges. The temporal evolution spans 6 years, with each time step representing one year. The dataset exhibits clear research community structures that evolve as researchers change their collaboration patterns and research interests.

\textbf{Brain Dataset:} This dataset represents functional brain networks derived from fMRI data, where nodes correspond to brain regions and edges represent functional connectivity between regions. The network contains 264 nodes (brain regions) and evolves over 200 time steps, each representing a different time window during a cognitive task. This dataset is particularly challenging due to its high-dimensional nature and complex temporal dynamics.

\begin{table}[ht]
\centering
\begin{tabular}{@{}lcccc@{}}
\toprule
Dataset & Nodes & Time Steps & Avg. Edges/Step & Domain \\
\midrule
LFR1-8 & 1,000 & 9 & $\sim$7,500 & Synthetic \\
PS & 242 & 125 & $\sim$400 & Social \\
CSMN & 1,899 & 193 & $\sim$310 & Social \\
DBLP & 12,334 & 6 & $\sim$25,000 & Collaboration \\
Brain & 264 & 200 & $\sim$1,200 & Biological \\
\bottomrule
\end{tabular}
\caption{ Statistics of the dynamic network datasets used in our experiments. Each dataset is characterized by the number of nodes, time steps, average edges per step, and its corresponding domain.}
\label{tab:datasets}
\end{table}

\subsection{Comparison Algorithms}

In order to evaluate the ability of the algorithm for community discovery in this article, it is mainly compared with dynamic community discovery algorithms Dyperm and FacetNet, static community discovery algorithm Louvain, dynamic community discovery method DyCTWE based on time walk graph embedding, and STEC-Net-K using K-means as the final clustering method. This algorithm is referred to as STEC-Net in this paper, and purity and NMI are used to evaluate the quality of clustering results.

The evaluation of spatiotemporal graph embedding models is performed by comparing them with traditional static graph embedding methods SDNE\citep{Wang2023}, DeepWalk\citep{Perozzi2014} and dynamic graph embedding algorithms DynAE, DynGEM\citep{Goyal2020}. After obtaining node embeddings, K-means clustering is used to divide the nodes into communities. In this paper, purity and normalized mutual information (NMI) are used to evaluate the quality of the clustering results. The parameter settings for the relevant dynamic graph embedding algorithm are shown in Tab.~\ref{tab:2}.

\begin{table}[ht]
    \centering
    \begin{tabular}{c|c}
    \hline \text { Algorithm } & \text { Parameter settings } \\
    \hline \text { DynAE } & $\beta=5, l b=2$ \\
    \hline \text { DynGEM } & $\beta=5, a=1e{-}5, v_1=v_2=1e{-}6, $ \\  
    & $\rho=0.3, \theta=0.01$ \\  
    \hline\text { DeepWalk } & $l=40, w=10, \gamma=6$ \\
    \hline
    \end{tabular}
    \caption{Parameter settings of related algorithms.}
    \label{tab:2}
\end{table}

To evaluate the community discovery capability of the proposed algorithm, we compared the STEC-Net-K, where K stands for the final K-means clustering, with dynamic community discovery algorithms Dyperm and FacetNet\citep{Lin2023}, the static community discovery algorithm Louvain\citep{Blondel2023}, and the dynamic community discovery method DyCTWE based on time walk graph embedding.

To further benchmark STEC-Net against attention-based approaches, we additionally compare with 
	ASTGCN~\cite{guo2019attention}, MSTGCN~\cite{jia2021multi}, and 
	ST-GAT~\cite{song2022st}, as well as the lightweight baseline 
	NWFCM~\cite{huang2024community}.
\subsection{Results}

\subsubsection{Comparison with Graph Embedding Methods}

\begin{figure}[ht]
\centering
\begin{subfigure}{0.49\linewidth}
    \includegraphics[width=\linewidth]{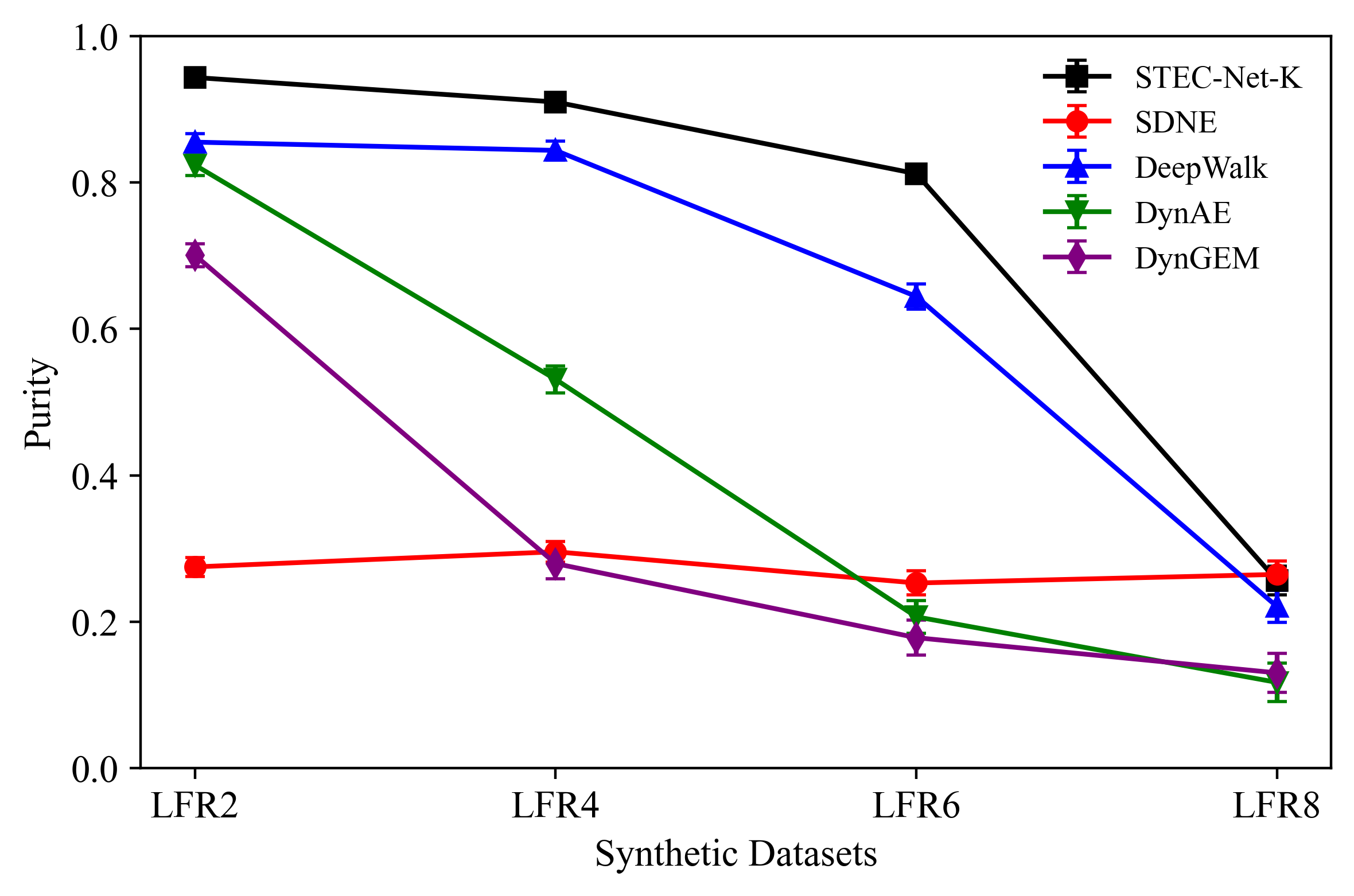}
    \caption{Average purity on Synthetic Datasets}
    \label{fig:4a}
\end{subfigure}
\begin{subfigure}{0.49\linewidth}
    \includegraphics[width=\linewidth]{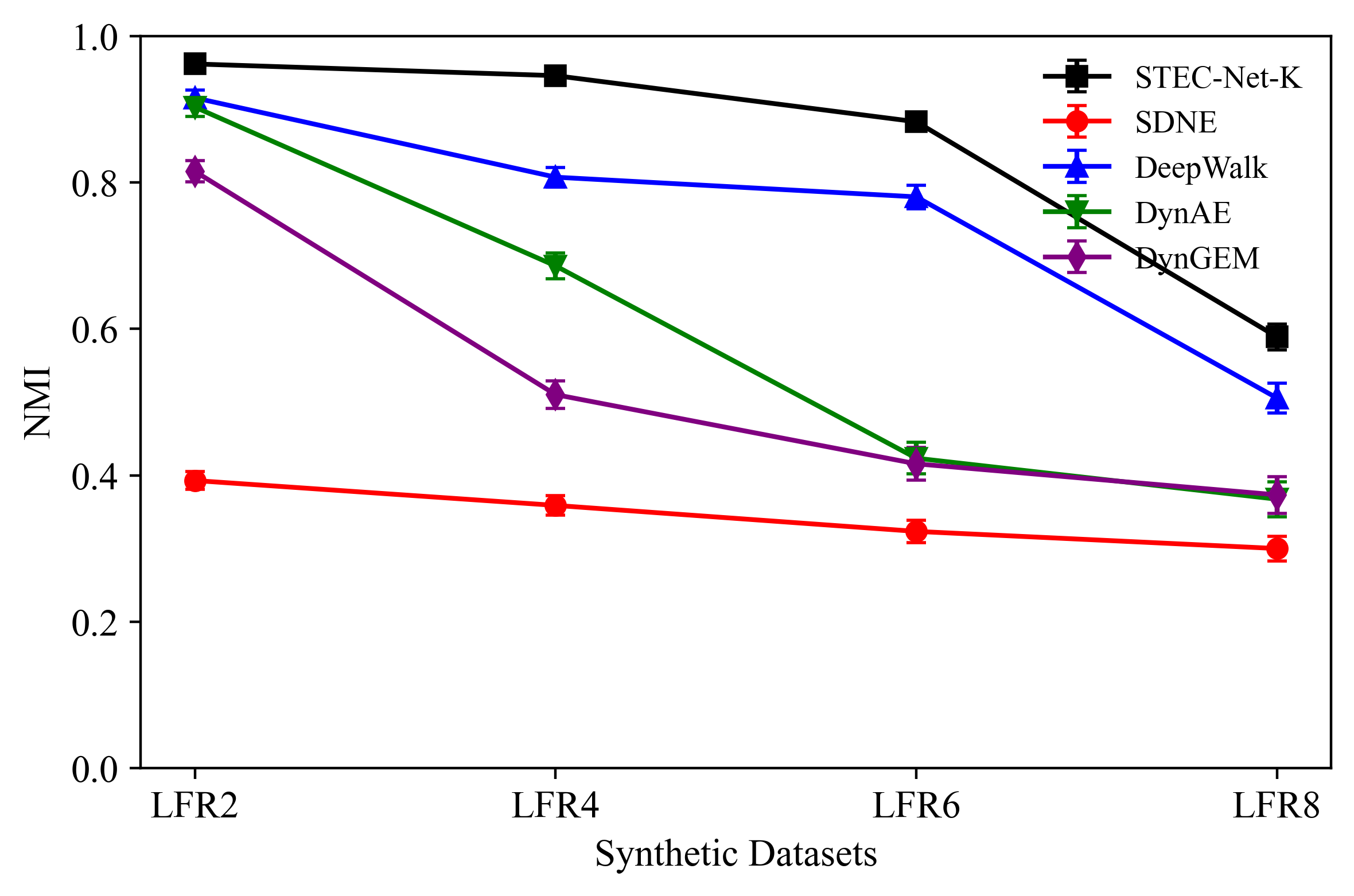}
    \caption{Average NMI on Synthetic Datasets}
    \label{fig:4b}
\end{subfigure}
\begin{subfigure}{0.49\linewidth}
    \includegraphics[width=\linewidth]{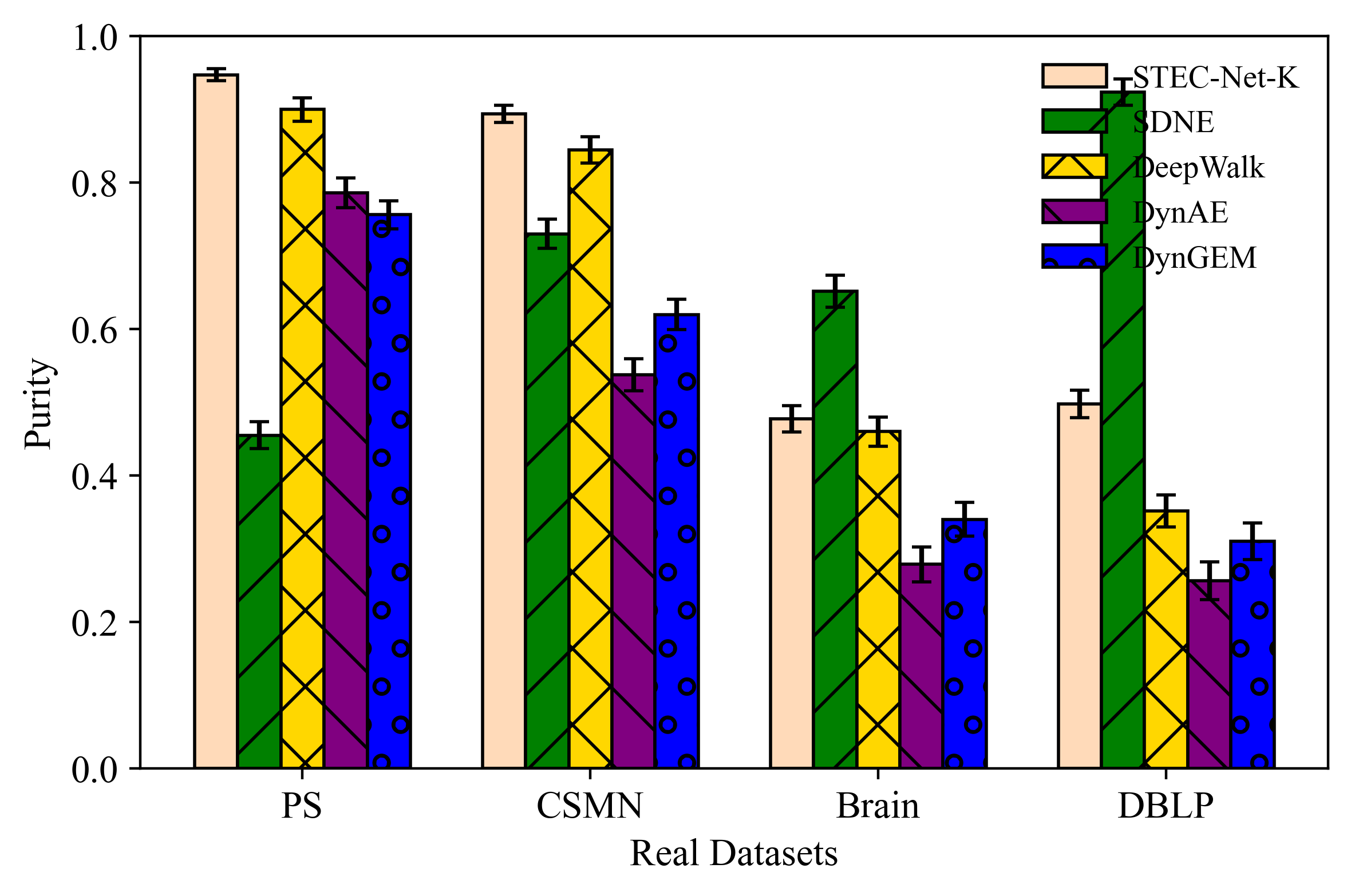}
    \caption{Average Purity on Real Datasets}
    \label{fig:4c}
\end{subfigure}
\begin{subfigure}{0.49\linewidth}
    \includegraphics[width=\linewidth]{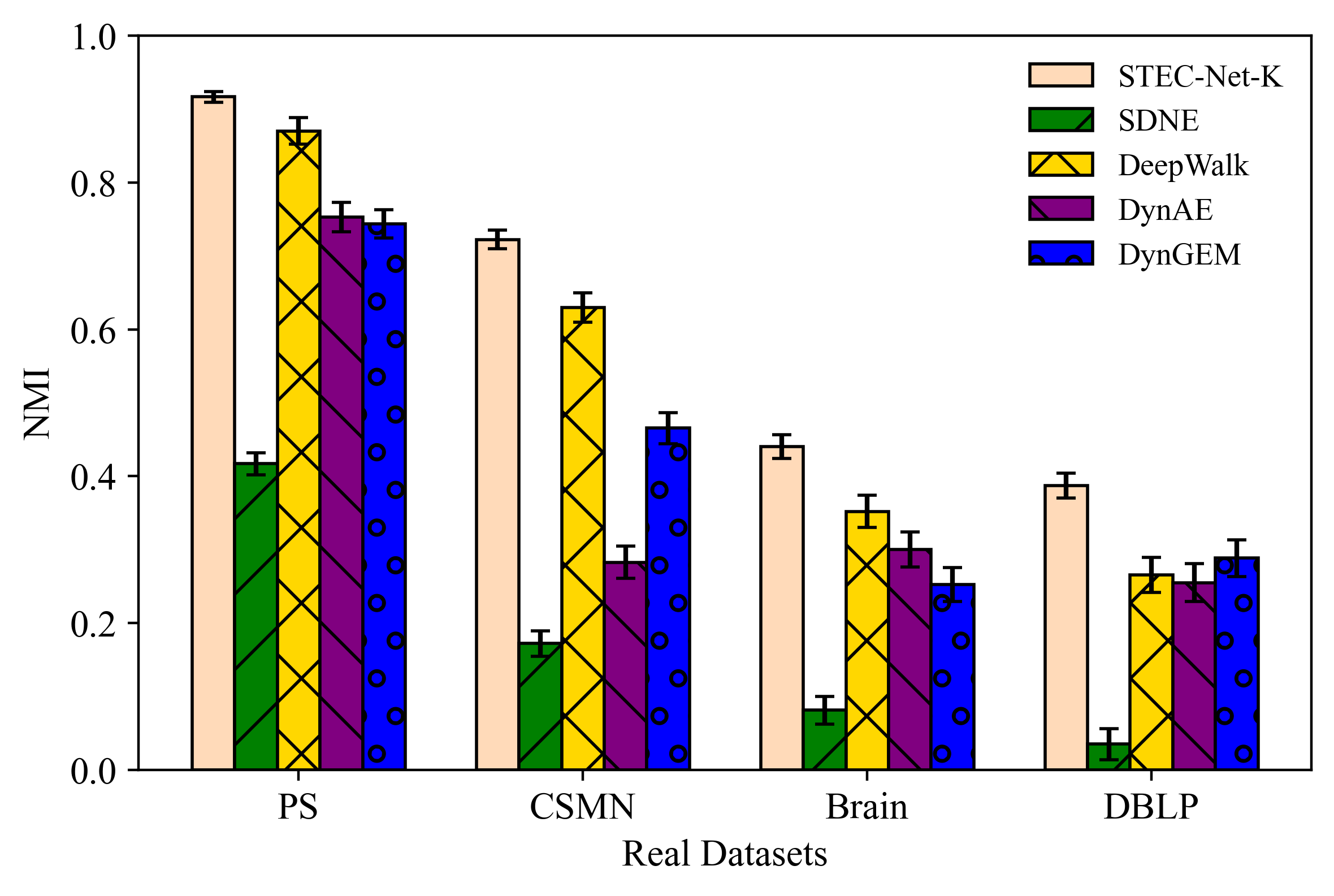}
    \caption{Average NMI on Real Datasets}
    \label{fig:4d}
\end{subfigure}
\caption{Performance comparison of graph embedding methods on synthetic and real datasets. Results show that STEC-Net consistently achieves higher purity and NMI scores compared to static methods (SDNE, DeepWalk) and dynamic methods (DynAE, DynGEM).}
\label{fig:4}
\end{figure}

In Fig.~\ref{fig:4} we present the evaluation results of various graph embedding algorithms on dynamic artificial synthetic datasets and real datasets. We first analyze the performance of the embedding algorithms on synthetic networks.

From Fig.~\ref{fig:4a} and ~\ref{fig:4b}, it is observed that as the complexity coefficient increases, the average evaluation results for synthetic networks show a decreasing trend. This indicates that as the complexity of the network grows, the ability of the graph embedding algorithm to represent the network structure declines. The STEC-Net-K algorithm exhibits strong representation capabilities for artificial networks, with K-means yielding superior community partitioning accuracy measured in purity and NMI compared to other embedding models. This demonstrates that the spatiotemporal graph embedding model effectively captures spatiotemporal relationships and preserves node interaction information in complex dynamic networks. The dynamic graph embedding methods DynGEM and DynAE can also provide good embedding features, and the evaluation results of community partitioning achieve good results when the network complexity is low. However, as complexity increases, the evaluation results rapidly decline. This may be attributed to the fact that these two algorithms primarily focus on learning the spatial structure of nodes during graph embedding. For instance, DynGEM only considers the influence of the previous time snapshot on the current one, neglecting the long-range dependencies of dynamic nodes. Consequently, as network complexity increases, the graph embedding performance declines, impacting community discovery effectiveness. The static embedding algorithm DeepWalk achieves strong results, while SDNE shows stable but modest performance on artificial datasets, with minimal changes in NMI values. This suggests that SDNE can consistently characterize dynamic artificial networks, but its ability to extract node features requires further improvement.

We then analyze the performance of the embedded model in real networks and present the results in Fig.~\ref{fig:4c} and ~\ref{fig:4d}.

From Fig.~\ref{fig:4c} and ~\ref{fig:4d}, it is evident that all graph embedding models perform better in node partitioning accuracy measured in purity and NMI on smaller dynamic datasets than on larger ones. The decrease in purity and NMI as the network size grows can be attributed to the increase in the number of nodes and the more complex interaction relationships between them. Larger networks introduce additional challenges, such as denser connections, overlapping communities, and a greater diversity of dynamic interactions, which make it harder for models to effectively capture and preserve all relevant spatiotemporal features.

Despite these challenges, the STEC-Net proposed in this paper consistently maintains high accuracy measured in purity and NMI across both small and large datasets. This highlights its robustness and its ability to effectively extract dynamic spatiotemporal information, allowing it to characterize the evolution of real-world networks more accurately. The model's ability to integrate spatial and temporal embeddings enables it to uncover subtle patterns and interactions that are often missed by other methods.

Among the baseline models, DeepWalk achieves the second-best performance due to its ability to learn meaningful node representations through random walks. However, its lack of temporal modeling limits its performance on dynamic datasets. SDNE, on the other hand, yields relatively lower purity and NMI on large-scale dynamic datasets, probably because its emphasis on preserving local structures struggles to scale effectively with the increasing complexity of larger networks. Among dynamic graph embedding algorithms, DynGEM outperforms DynAE in community discovery, highlighting that SDNE's embedding performance on the network surpasses that of fully connected neural networks.

\subsubsection{Comparison with Community Discovery Methods}

\begin{figure}[ht]
\centering
\begin{subfigure}{0.49\linewidth}
    \includegraphics[width=\linewidth]{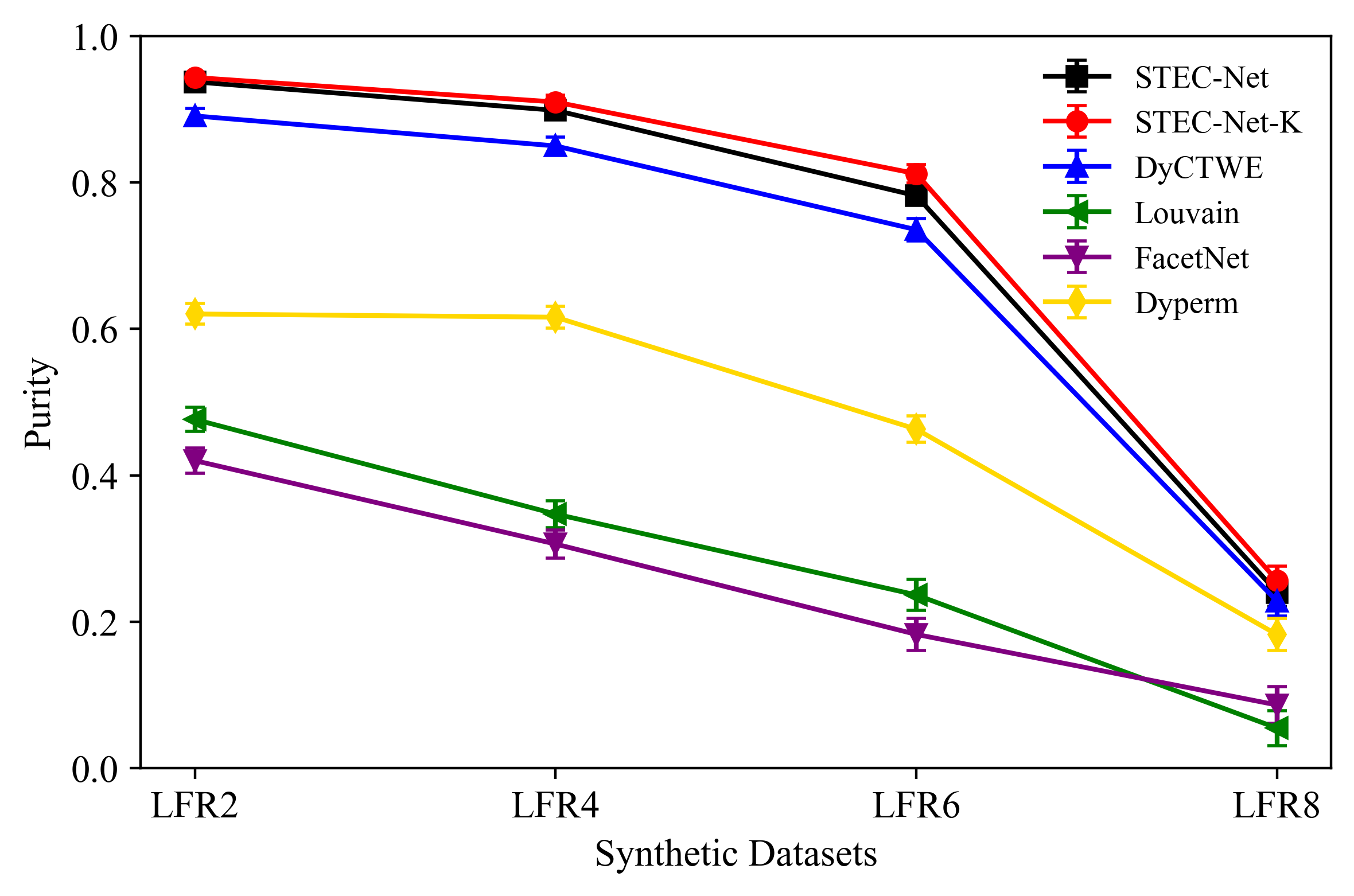}
    \caption{Average purity on Synthetic Datasets}
    \label{fig:5a}
\end{subfigure}
\begin{subfigure}{0.49\linewidth}
    \includegraphics[width=\linewidth]{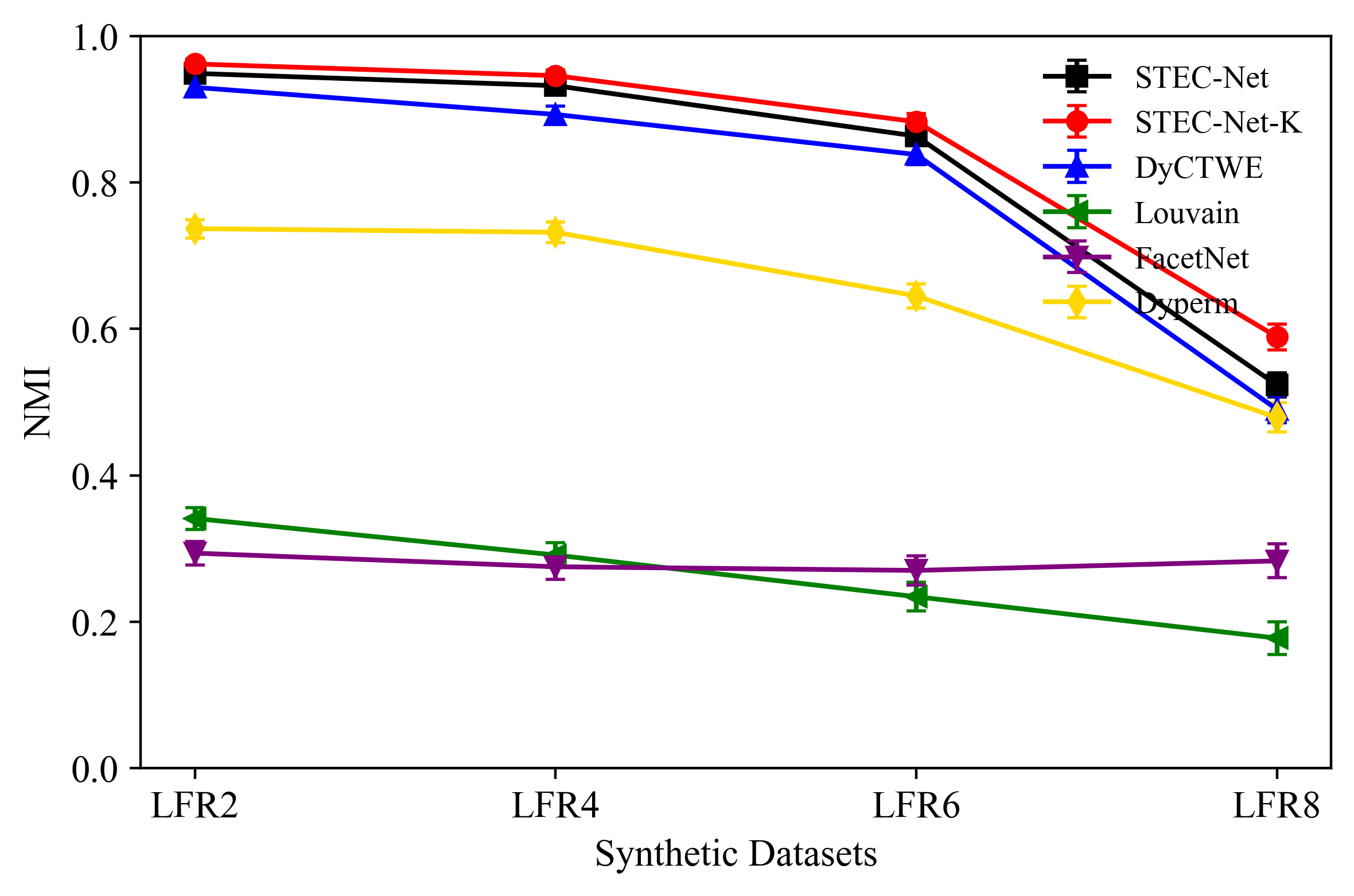}
    \caption{Average NMI on Synthetic Datasets}
    \label{fig:5b}
\end{subfigure}
\begin{subfigure}{0.49\linewidth}
    \includegraphics[width=\linewidth]{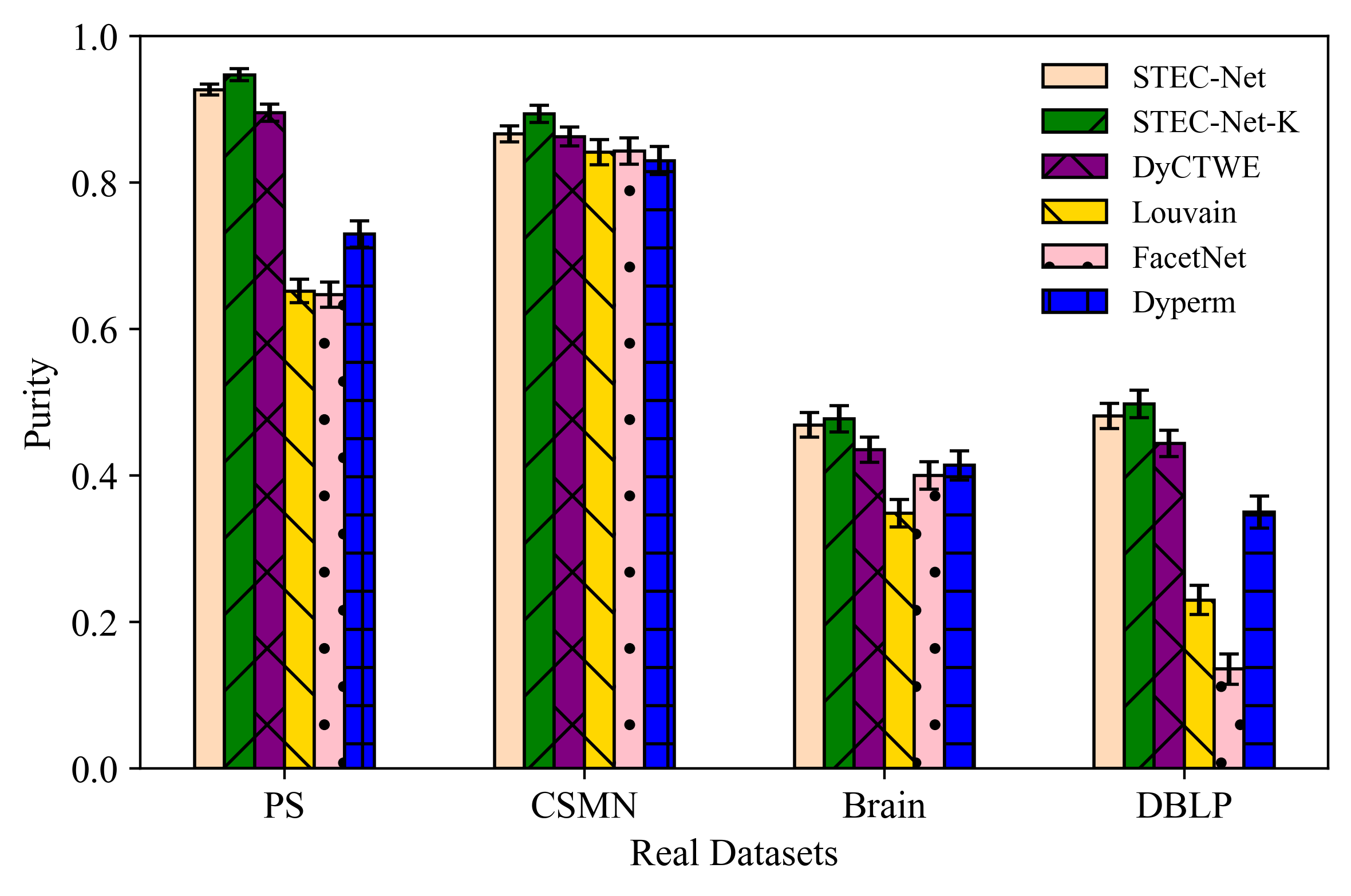}
    \caption{Average Purity on Real Datasets}
    \label{fig:5c}
\end{subfigure}
\begin{subfigure}{0.49\linewidth}
    \includegraphics[width=\linewidth]{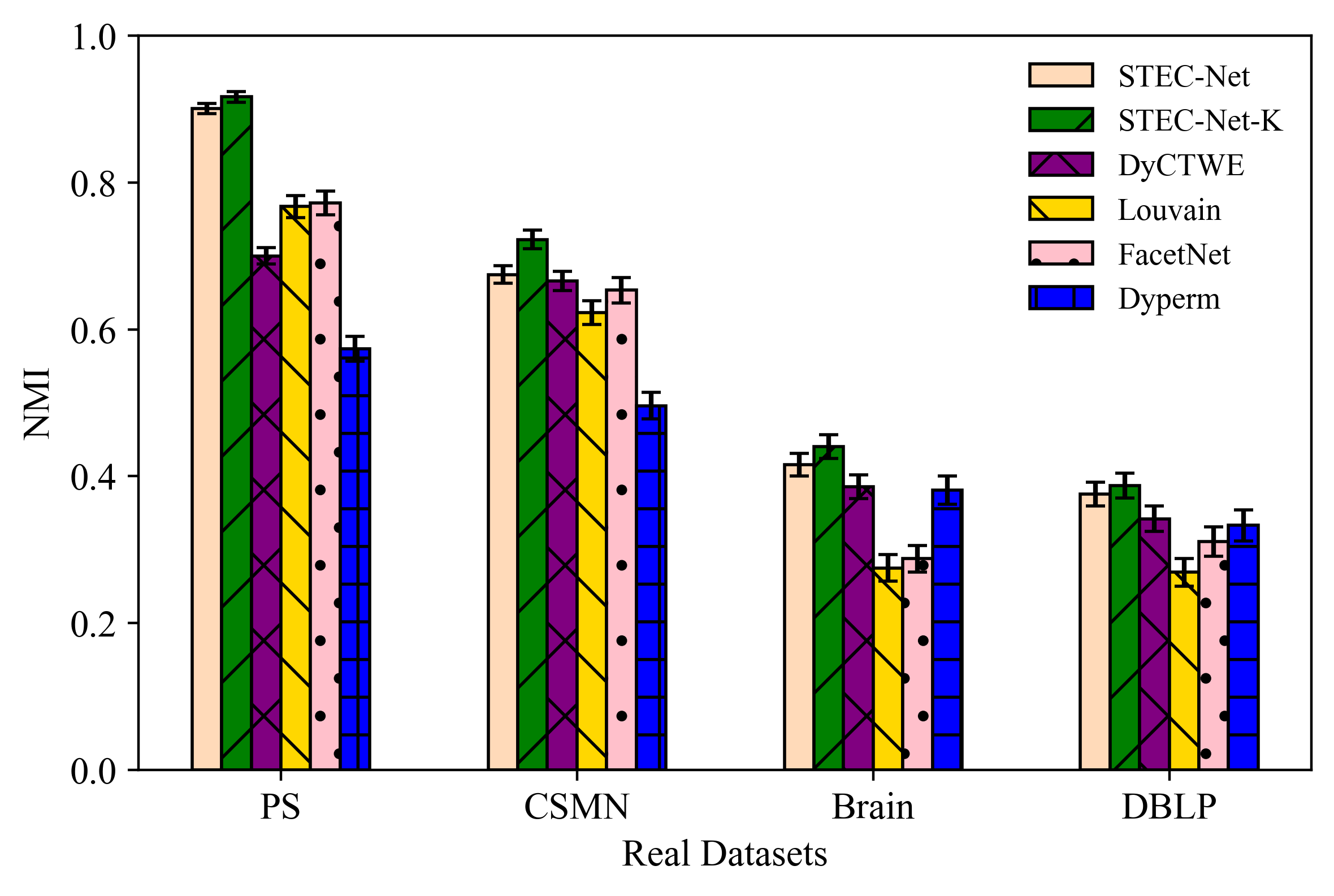}
    \caption{Average NMI on Real Datasets}
    \label{fig:5d}
\end{subfigure}
\caption{Comprehensive comparison of community discovery algorithms. STEC-Net demonstrates superior performance over traditional static (Louvain), dynamic (Dyperm, FacetNet), and embedding-based (DyCTWE) methods across both synthetic and real datasets.}
\label{fig:5}
\end{figure}

In Fig.~\ref{fig:5a} and ~\ref{fig:5d} we present the evaluation results of community discovery algorithms on dynamic networks. From these figures, STEC-Net demonstrates strong  community partitioning performance across both synthetic and real-world dynamic network datasets. The evaluation metrics, purity and NMI across all datasets, are presented in Tab.~\ref{tab:accuracy}. In general, the variability is lower on easier synthetic datasets and increases on more complex or ambiguous datasets, especially LFR8, Brain, and DBLP, which is consistent with the greater difficulty of stable community identification in these settings.

\begin{figure}[t]
	\centering
	\includegraphics[width=0.8\linewidth]{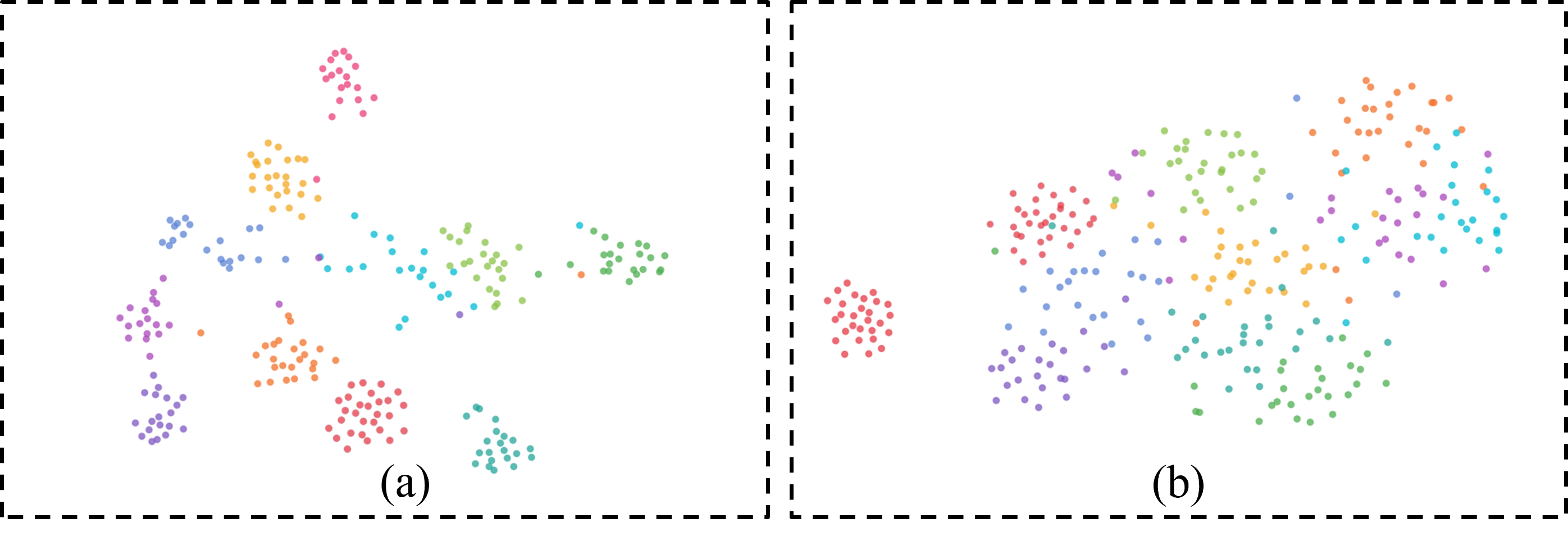}
	\caption{t-SNE visualization of node embeddings learned 
		by STEC-Net at a representative time step. 
		(a) PS dataset: well-separated clusters indicate 
		clearly distinguishable community structure. 
		(b) Brain dataset: heavily overlapping embeddings 
		reflect the inherently ambiguous and non-stationary 
		community structure of functional brain networks.}
	\label{fig:tsne}
\end{figure}

However, performance degrades notably on larger datasets, 
particularly DBLP (Purity: 0.4813, NMI: 0.3758) and Brain 
(Purity: 0.4689, NMI: 0.4158). As shown in 
Figure~\ref{fig:5c} and \ref{fig:5d}, all baseline 
methods also suffer significant degradation on these datasets, 
suggesting the difficulty stems from intrinsic dataset 
characteristics rather than limitations specific to STEC-Net. 
For DBLP, the sparse temporal resolution (only 6 time steps) 
combined with a high node count (12,334 nodes) limits the 
GRU's ability to capture long-range dependencies. For the 
Brain dataset, highly non-stationary functional connectivity 
across 200 time steps prevents stable community formation, 
as evidenced by the diffuse embedding distribution in 
figure~\ref{fig:tsne}(b), in contrast to the well-separated 
clusters observed for the PS dataset 
in figure~\ref{fig:tsne}(a). These findings highlight 
scalability and soft community assignment as key directions 
for future work.

\begin{table}[ht]
	\centering
	\begin{tabular}{@{}lcc@{}}
		\toprule
		\textbf{Dataset} & \textbf{Purity} & \textbf{NMI} \\ \midrule
		LFR2  & \textcolor{black}{$0.9375 \pm 0.0068$} & \textcolor{black}{$0.9491 \pm 0.0059$} \\
		LFR4  & \textcolor{black}{$0.8986 \pm 0.0084$} & \textcolor{black}{$0.9323 \pm 0.0071$} \\
		LFR6  & \textcolor{black}{$0.7819 \pm 0.0116$} & \textcolor{black}{$0.8634 \pm 0.0102$} \\
		LFR8  & \textcolor{black}{$0.2401 \pm 0.0189$} & \textcolor{black}{$0.5233 \pm 0.0164$} \\
		PS    & \textcolor{black}{$0.9270 \pm 0.0075$} & \textcolor{black}{$0.9008 \pm 0.0068$} \\
		CSMN  & \textcolor{black}{$0.8666 \pm 0.0107$} & \textcolor{black}{$0.6747 \pm 0.0121$} \\
		Brain & \textcolor{black}{$0.4689 \pm 0.0168$} & \textcolor{black}{$0.4158 \pm 0.0153$} \\
		DBLP  & \textcolor{black}{$0.4813 \pm 0.0175$} & \textcolor{black}{$0.3758 \pm 0.0161$} \\
		\bottomrule
	\end{tabular}
	\caption{Purity and NMI in the synthetic and real datasets of the proposed STEC-Net algorithm.}
	\label{tab:accuracy}
\end{table}

Comparing the proposed algorithm with traditional static community discovery algorithms, such as Louvain, and dynamic community discovery algorithms such as Dyperm and FacetNet, STEC-Net outperforms these methods in community partitioning. FacetNet's performance on artificial networks is suboptimal, although it performs better on real networks, suggesting that it focuses more on changes in real-world social activities. Dyperm demonstrates relatively stable community partitioning performance, while Louvain performs well in real online communities.

\begin{figure}[ht]
\centering
\includegraphics[width=0.5\linewidth]{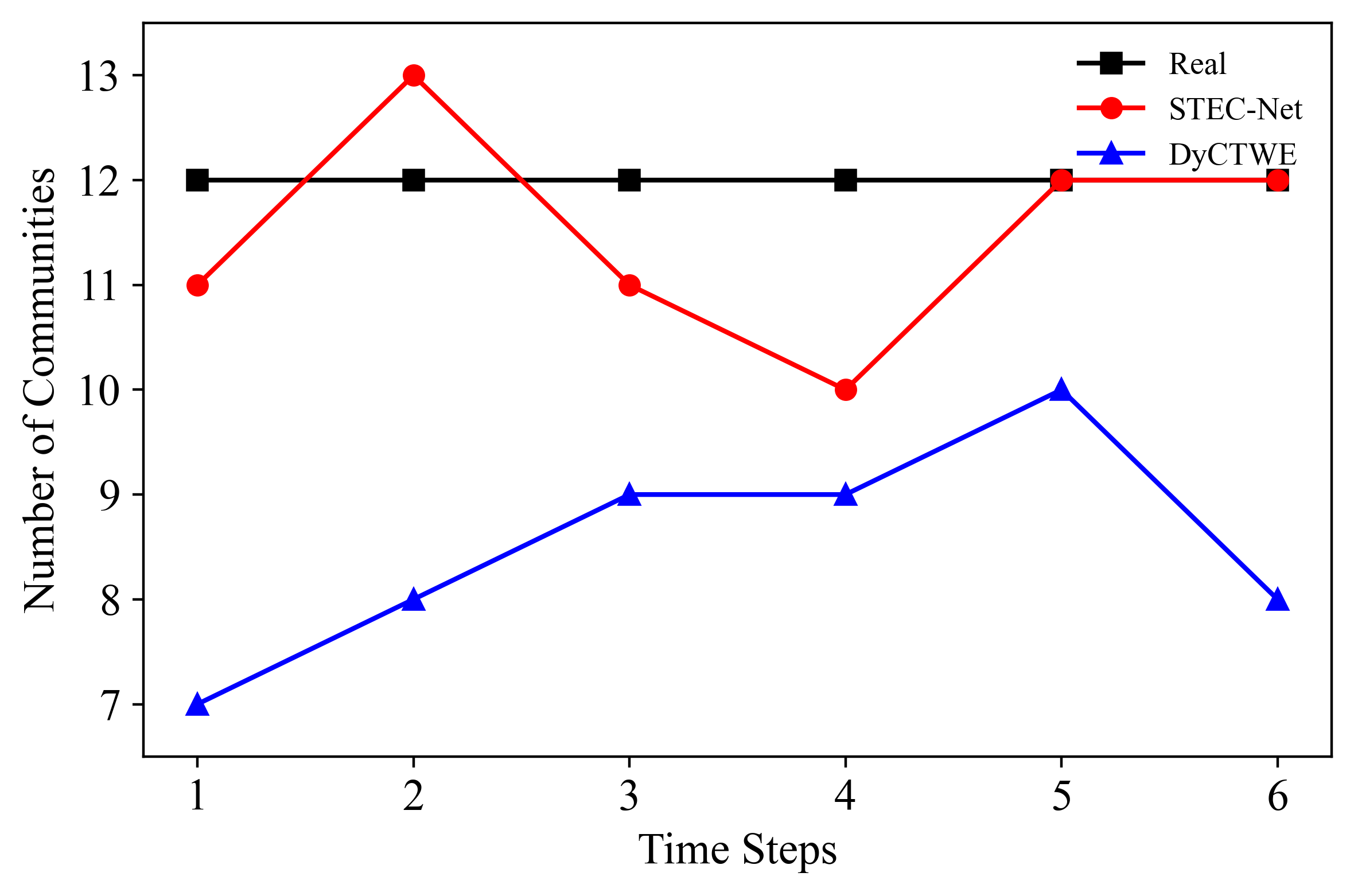}
\caption{Number of communities identified by STEC-Net and DyCTWE on the PS dataset, compared with the ground-truth community structure across time steps.}
\label{fig:8}
\end{figure}

The comparison between STEC-Net and STEC-Net-K primarily evaluates the impact of clustering algorithms on the partitioning results of the graph embedding model. From the results, it is observed that the evaluation results for both the SOM and the K-means clustering algorithms are similar, with differences within 5\%. The SOM algorithm was chosen because it does not require specifying the number of clusters in advance, only the number of competing units, neighborhood range, and learning rate. This allows for adaptive community division based on the embedding representation of nodes. In contrast, the K-means algorithm requires prior knowledge of the number of clusters, which is often difficult to determine in real dynamic networks. Hence, SOM clustering is more suitable for this work.

\begin{figure}[ht]
\centering
\begin{subfigure}{\linewidth}
    \includegraphics[width=0.5\linewidth]{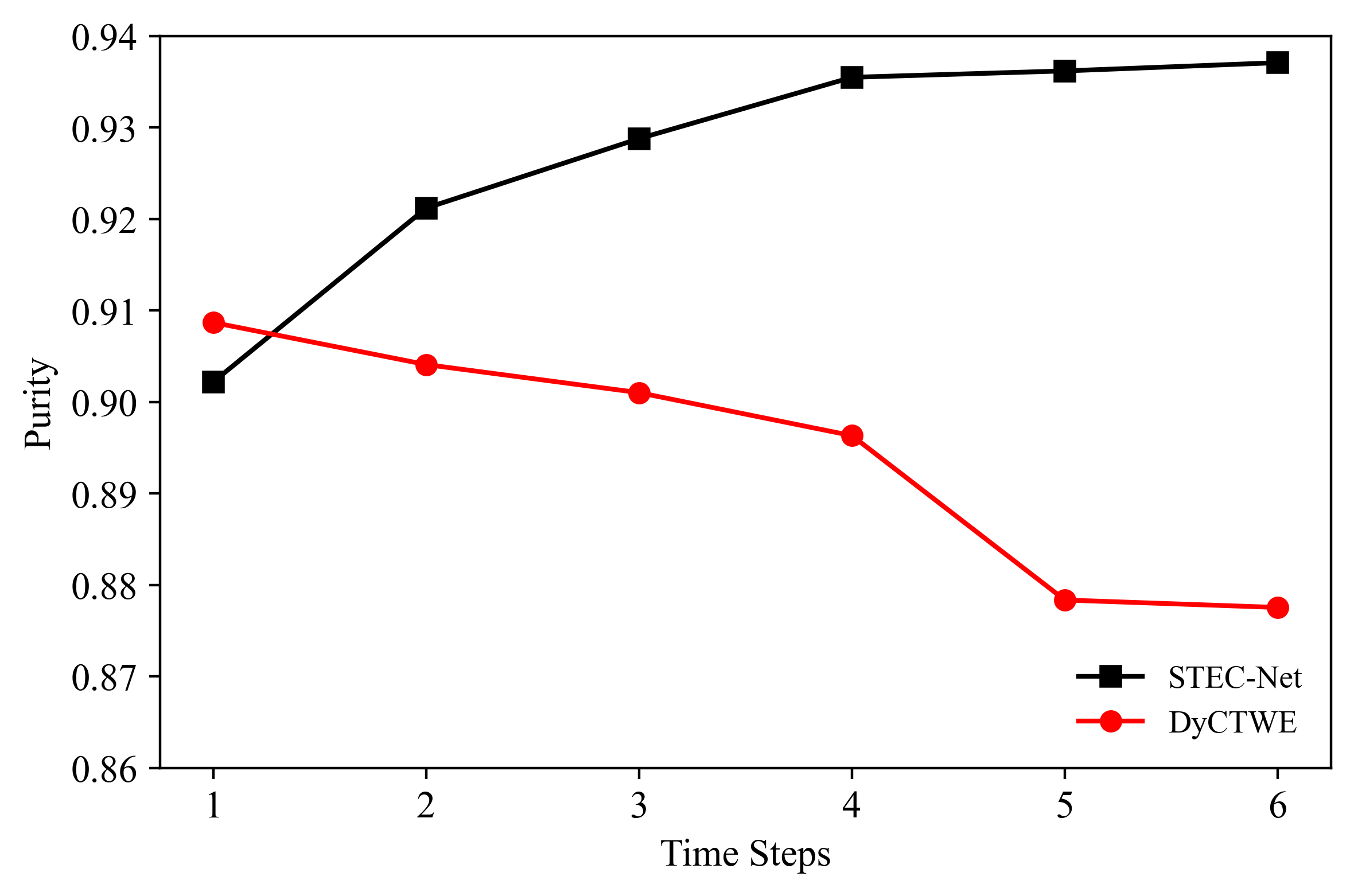}
    \caption{Average purity in PS}
    \label{fig:9a}
\end{subfigure}
\begin{subfigure}{\linewidth}
    \includegraphics[width=0.5\linewidth]{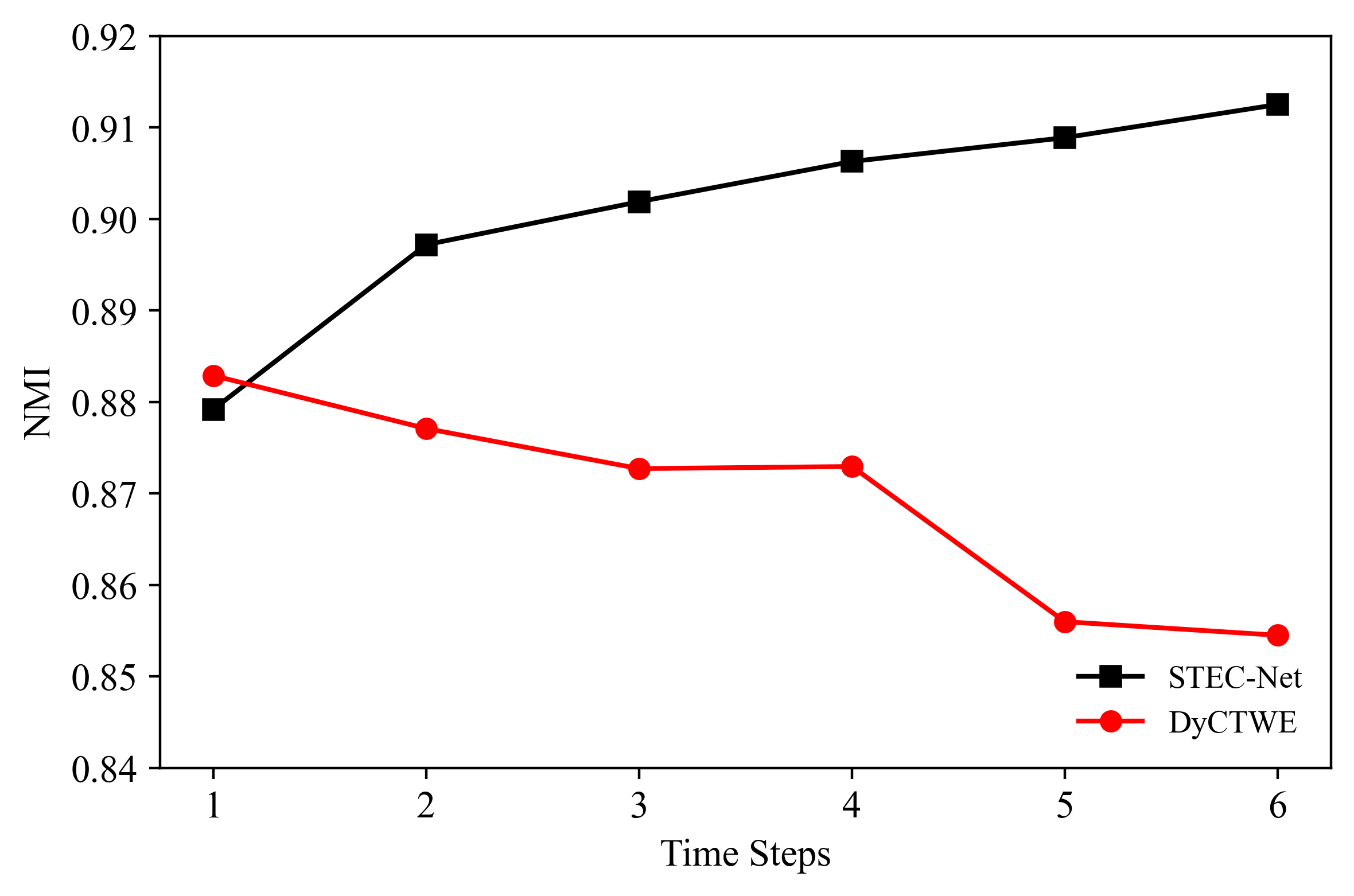}
    \caption{Average NMI in PS}
    \label{fig:9b}
\end{subfigure}
\caption{Evaluation results of STEC-Net and DyCTWE on the PS dataset over time. The plots report Purity (left) and NMI (right) at each time step, showing that STEC-Net maintains consistently higher performance across the dynamic snapshots.}
\label{fig:9}
\end{figure}

In Fig.~\ref{fig:8}, we present the comparison of community partitioning results for the PS dataset, while in Fig.~\ref{fig:9}, we present the evaluation results of both algorithms on the same dataset. From Fig.~\ref{fig:8}, it is evident that STEC-Net's partitioning results are closely aligned with the actual number of communities, with only minor discrepancies. In contrast, DyCTWE shows a maximum difference of 5 communities from the real partitioning. Fig.~\ref{fig:9} reveals that the quality of DyCTWE's partitioning deteriorates over time, indicating error accumulation and a decline in community discovery performance. This issue likely arises because DyCTWE overly emphasizes local incremental updates, neglecting the overall evolution of the network. On the other hand, STEC-Net demonstrates an increasing trend in performance over time, suggesting that spatiotemporal graph embedding effectively captures the dynamic evolution of nodes and offers robust representation capabilities for dynamic networks.

\subsubsection{\textcolor{black}{Comparison with Attention-Based Methods}}

\textcolor{black}{In this section, we compare STEC-Net against the following 
	methods in terms of purity, NMI, and computational cost:}

\begin{itemize}
	
	\item \textcolor{black}{\textbf{ASTGCN}: 
		An attention-based spatial-temporal graph convolutional network 
		originally proposed for traffic flow forecasting.}
	
	\item \textcolor{black}{\textbf{MSTGCN}: 
		A multi-view spatial-temporal graph convolutional network using 
		attention mechanisms.}
	
	\item \textcolor{black}{\textbf{ST-GAT}: 
		A spatio-temporal graph attention network for spatiotemporal 
		modeling.}
		
		\item \textcolor{black}{\textbf{NWFCM}: 
		A node-weighted fuzzy c-means method that incorporates node importance 
		into the clustering objective, serving as a lightweight traditional 
		baseline.}
	
\end{itemize}

The results in Table~\ref{tab:comprehensive_comparison} demonstrate that the proposed STEC-Net algorithm exhibits outstanding performance in both community detection quality and computational efficiency across two representative datasets: LFR2 and PS.

\begin{widetext}
\begin{table}[ht]
	\centering
	\caption{Comprehensive Performance Comparison of Different Algorithms.}
	\begin{tabular}{@{}lccccr@{}}
		\toprule
		\multirow{2}{*}{Algorithm} & \multirow{2}{*}{Dataset} & \multirow{2}{*}{Purity} & \multirow{2}{*}{NMI} & Training Time & Memory/Params \\
		&  &  &  & (min) &  \\
		\midrule
		\multirow{2}{*}{STEC-Net}
		& LFR2 & \textbf{\textcolor{black}{$0.9375 \pm 0.0068$}} & \textbf{\textcolor{black}{$0.9491 \pm 0.0059$}} & 12.3 & 2.1GB/48K \\
		& PS   & \textbf{\textcolor{black}{$0.9270 \pm 0.0075$}} & \textbf{\textcolor{black}{$0.9008 \pm 0.0068$}} & 8.5  & 1.8GB/48K \\
		\multirow{2}{*}{NWFCM}
		& LFR2 & \textcolor{black}{$0.7203 \pm 0.0214$} & \textcolor{black}{$0.6974 \pm 0.0230$} & 6.2  & 0.8GB/12K \\
		& PS   & \textcolor{black}{$0.7047 \pm 0.0241$} & \textcolor{black}{$0.6823 \pm 0.0255$} & 4.1  & 0.6GB/12K \\
		\multirow{2}{*}{ASTGCN}
		& LFR2 & \textcolor{black}{$0.8353 \pm 0.0198$} & \textcolor{black}{$0.8218 \pm 0.0211$} & 28.6 & 4.5GB/156K \\
		& PS   & \textcolor{black}{$0.7972 \pm 0.0245$} & \textcolor{black}{$0.7857 \pm 0.0227$} & 18.7 & 3.8GB/156K \\
		\multirow{2}{*}{MSTGCN}
		& LFR2 & \textcolor{black}{$0.8571 \pm 0.0186$} & \textcolor{black}{$0.8434 \pm 0.0199$} & 45.3 & 6.8GB/243K \\
		& PS   & \textcolor{black}{$0.8403 \pm 0.0228$} & \textcolor{black}{$0.8281 \pm 0.0214$} & 32.1 & 5.9GB/243K \\
		\multirow{2}{*}{ST-GAT}
		& LFR2 & \textcolor{black}{$0.8127 \pm 0.0222$} & \textcolor{black}{$0.7943 \pm 0.0204$} & 22.4 & 3.7GB/128K \\
		& PS   & \textcolor{black}{$0.7896 \pm 0.0251$} & \textcolor{black}{$0.7738 \pm 0.0236$} & 16.2 & 3.1GB/128K \\
		\bottomrule
	\end{tabular}
	\label{tab:comprehensive_comparison}
\end{table}
\end{widetext}

The proposed STEC-Net achieves Purity values of 0.9375 and 0.9270 on the LFR2 and PS datasets, respectively, with corresponding NMI values of 0.9491 and 0.9008, significantly outperforming all baseline algorithms. Compared to the second-best performing MSTGCN, STEC-Net shows an average improvement of approximately 8.2\% in Purity and 9.1\% in NMI. This indicates that STEC-Net, through its spatiotemporal joint modeling strategy combining GCN and GRU, can more accurately capture community evolution patterns in dynamic networks than attention-based approaches. 

The performance comparison reveals distinct characteristics of different algorithm categories:
\begin{itemize}
	\item \textbf{Traditional methods} like NWFCM show the lowest performance but maintain computational efficiency.
	\item \textbf{Attention-based methods} (ASTGCN, MSTGCN, ST-GAT) achieve moderate performance but with significant computational overhead.
	\item \textbf{Our spatiotemporal approach} achieves the best performance while maintaining reasonable computational cost.
\end{itemize}

STEC-Net also demonstrates clear efficiency advantages over attention-based methods. Although its training time is slightly higher than that of the traditional NWFCM approach, it remains substantially lower than that of more complex attention-based models. Specifically, compared with MSTGCN, STEC-Net reduces training time by approximately 70\% and memory consumption by about 65\%. Relative to ASTGCN, it achieves 57\% faster training and 53\% lower memory usage. In comparison with ST-GAT, STEC-Net attains a 45\% reduction in training time and a 43\% decrease in memory consumption. Additionally, STEC-Net requires only 48K parameters, significantly fewer than ASTGCN's 156K, MSTGCN's 243K, and ST-GAT's 128K parameters, reflecting the simplicity and efficiency of the model design compared to attention mechanisms that require substantial parameter overhead.

The robustness of STEC-Net is further demonstrated by its stable performance across datasets of different nature and complexity. From the synthetic LFR2 to the real-world PS network, the decline in Purity is less than 2\%, while attention-based algorithms typically experience 3-8\% decreases. This demonstrates the superior generalization capability and stability of the proposed algorithm. The consistent performance across different dataset characteristics (synthetic vs. real-world, different scales and domains) indicates that our spatiotemporal modeling approach is more robust than attention mechanisms that may be sensitive to specific network structures.

The results further underscore the advantages of STEC-Net over attention-based methods. Specifically, ASTGCN, although effective for traffic prediction, shows limited suitability for community detection tasks. MSTGCN leverages multi-view attention mechanisms but incurs substantial computational overhead, making it less practical for large or complex networks. ST-GAT integrates graph attention networks for spatiotemporal modeling, yet its performance exhibits instability across different datasets. In conclusion, the proposed STEC-Net achieves superior detection accuracy measured in purity and NMI, and computational efficiency, offering more stable and effective spatiotemporal modeling than attention-based counterparts, while maintaining substantially lower computational complexity. \textcolor{black}{Furthermore, the performance gaps between STEC-Net and all baselines consistently exceed three standard deviations across all repeated experiments, indicating that the observed improvements are not attributable to random variation.}

\subsubsection{\textcolor{black}{Ablation Study}}
In complex machine learning models, which often consist of multiple modules, ablation experiments are used to assess the importance of specific components by removing certain modules and rerunning the experiments. The results are then compared with those from the original model to evaluate the impact of the removed modules. We evaluate the effectiveness of dynamic community methods based on spatiotemporal graph embedding through ablation experiment. This experiment is conducted on a dynamic artificial network (LFR2) with $\mu=0.2$. We compare the full model with GRU, GCNs, and GCNs-GRU. The evaluation criteria include the average NMI, completeness, and homogeneity of the dynamic network, which are used to verify the superiority of the proposed algorithm. A detailed comparison is presented in Tab~\ref{tab:ab}.

As shown in Tab.~\ref{tab:ab}, the proposed STEC-Net consistently outperforms its individual components, demonstrating its effectiveness in capturing the complex spatiotemporal relationships of dynamic networks. STEC-Net achieves superior performance across all evaluation metrics, including NMI, completeness, and homogeneity, confirming that its integrated design is well-suited for dynamic community discovery. The relatively low performance of the GRU baseline can be attributed to its emphasis on temporal dynamics while neglecting spatial structural characteristics. Without explicit modeling of network topology, GRU struggles to capture node dependencies, leading to weaker community detection. In addition, the random initialization of node features without fixed semantics introduces inconsistencies in representation, further diminishing its effectiveness. Furthermore, the GCNs-GRU hybrid achieves substantially better results than GCNs alone, highlighting the importance of jointly modeling spatial topology and temporal evolution. This synergy between GCNs and GRU provides a more comprehensive representation of evolving communities and yields consistently higher performance across all metrics. \textcolor{black}{Furthermore, STEC-Net yields the smallest standard deviation 
	across all variants, indicating more consistent and reliable performance 
	across different random initializations. The differences between STEC-Net and all ablated variants also exceed three standard deviations, confirming the statistical reliability of each component's contribution.}

\begin{table}[ht]
	\centering
	\begin{tabular}{@{}lccc@{}}
		\toprule
		\textbf{Algorithm} & \textbf{NMI} & \textbf{Comp} & \textbf{Homo} \\ \midrule
		GCNs
		& \textcolor{black}{$0.7481 \pm 0.0126$}
		& \textcolor{black}{$0.7480 \pm 0.0135$}
		& \textcolor{black}{$0.7483 \pm 0.0122$} \\
		GRU
		& \textcolor{black}{$0.7055 \pm 0.0154$}
		& \textcolor{black}{$0.7054 \pm 0.0148$}
		& \textcolor{black}{$0.7056 \pm 0.0151$} \\
		GCNs-GRU
		& \textcolor{black}{$0.9019 \pm 0.0083$}
		& \textcolor{black}{$0.8999 \pm 0.0091$}
		& \textcolor{black}{$0.9021 \pm 0.0080$} \\
		STEC-Net
		& \textbf{\textcolor{black}{$0.9491 \pm 0.0059$}}
		& \textbf{\textcolor{black}{$0.9482 \pm 0.0063$}}
		& \textbf{\textcolor{black}{$0.9454 \pm 0.0066$}} \\
		\bottomrule
	\end{tabular}
	\caption{Comparison of ablation experimental results on LFR2 dataset ($\mu=0.2$). The table demonstrates the effectiveness of different component combinations in STEC-Net.}
	\label{tab:ab}
\end{table}

%\subsubsection{\textcolor{black}{Analysis of Performance Degradation 
%		on Large Datasets}}
%
%\textcolor{black}{The relatively low performance on DBLP and Brain 
%	datasets (Purity: 0.4813 and 0.4689, respectively) can be 
%	attributed to several factors:}
%
%\begin{itemize}
%	\item \textcolor{black}{\textbf{Sparse temporal snapshots}: 
%		DBLP contains only 6 time steps, which limits the GRU's 
%		ability to model long-range temporal dependencies effectively.}
%	
%	\item \textcolor{black}{\textbf{High node count with complex 
%			community overlap}: DBLP has 12,334 nodes with dense 
%		co-authorship relationships, leading to overlapping community 
%		structures that the current hard-partition SOM clustering 
%		cannot fully resolve.}
%	
%	\item \textcolor{black}{\textbf{Complex temporal dynamics}: 
%		The Brain dataset contains 200 fine-grained time steps with 
%		highly non-stationary functional connectivity, making it 
%		difficult for snapshot-level GCN to capture stable community 
%		structures.}
%\end{itemize}
%
%\textcolor{black}{These observations suggest that future work 
%	should explore overlapping community detection and adaptive 
%	temporal modeling for large-scale dynamic networks.}

\section{Conclusion}\label{sec:conc}

In this work, we proposed STEC-Net, a community discovery algorithm based on spatiotemporal graph embedding, to explore the complex structure and dynamics of social networks. Experimental validation demonstrates that STEC-Net effectively uncovers community evolution in dynamic networks, providing new insights into group organization and information flow.

A key characteristic of STEC-Net is its discrete-time snapshot framework, which is well suited to datasets collected in aggregated intervals. While continuous-time approaches such as TGN and JODIE offer fine-grained event-level modeling, our datasets are not inherently event-driven, and converting them into event streams could distort their temporal structure. For this reason, we adopt the snapshot-based representation, while extending STEC-Net to continuous-time settings with appropriate datasets remains an important direction for future research.

Looking ahead, we plan to enhance STEC-Net with advanced deep learning architectures, such as attention-based networks, to strengthen spatiotemporal feature learning and improve performance across evaluation metrics. We will also extend the framework to diverse domains, including transportation, biology, and particle interactions, to test its generality. Of particular significance is the financial sector, where applications in fraud detection and anti–money laundering can uncover hidden transactional patterns and anomalous group behaviors that traditional methods often miss. Such extensions would not only advance domain-specific impact but also demonstrate the broader practical utility of STEC-Net for understanding complex and evolving networks.

\section*{Acknowledgment}
The authors gratefully acknowledge the support of Qilu Bank, which provided the necessary CPU and GPU machine time for the computational experiments. We also wish to express our gratitude to its Financial Mathematical Lab for providing a conducive research environment and valuable resources.

\section*{Data Availability Statement}
This study utilized publicly available open datasets, including real-world dynamic network data and benchmark datasets. The synthetic datasets generated and analyzed during the study are available from the corresponding author upon reasonable request.

\bibliographystyle{JHEP}
\bibliography{bib_wiley}
\end{document}